\documentclass[5p,times,authoryear,nopreprintline]{elsarticle}

\usepackage{framed}
\usepackage{amsmath,amssymb,amsfonts}
\usepackage{graphicx}
\usepackage{textcomp}
\usepackage{booktabs}
\usepackage{multirow}
\usepackage{array}
\usepackage{tabularx}
\usepackage{url}
\usepackage{placeins}
\usepackage{flafter}
\usepackage{stfloats}
\usepackage{etoolbox}
\usepackage{xcolor}
\usepackage[hidelinks]{hyperref}
\hypersetup{pdfauthor={Shaojun Li, Xiaoyu Deng, Ikuo Keshi},pdftitle={TransNSNet: A Transformer-Enhanced Multi-Scale Adversarial Framework for CT-to-PET Image Synthesis}}

\DeclareUnicodeCharacter{0306}{}
\DeclareUnicodeCharacter{2212}{-}

\newcolumntype{Y}{>{\raggedright\arraybackslash}X}
\newcolumntype{C}{>{\centering\arraybackslash}X}

\begin{document}

\begin{frontmatter}

\title{TransNSNet: A Transformer-Enhanced Multi-Scale Adversarial Framework for CT-to-PET Image Synthesis}

\author[1]{Shaojun Li}
\author[2]{Xiaoyu Deng}
\author[1]{Ikuo Keshi\corref{cor1}}
\ead{keshi@fukui-ut.ac.jp}

\cortext[cor1]{Corresponding author.}

\address[1]{Fukui University of Technology, Fukui Campus, 3-6-1 Gakuen, Fukui City, Fukui Prefecture 910-8505, JAPAN}
\address[2]{University of Fukui, Bunkyo Campus, 3-9-1 Bunkyo, Fukui City, Fukui Prefecture 910-8507, JAPAN}

\begin{abstract}
This study investigates CT-to-PET image synthesis using TransNSNet, a Transformer-enhanced multi-scale adversarial framework. A paired CT-PET dataset was curated from the public non-small cell lung cancer (NSCLC) Radiogenomics collection through patient screening, slice-level matching, preprocessing, and patient-level train-test separation. The proposed framework generates PET-like predictions at multiple spatial scales and applies coarse-to-fine reconstruction supervision, while TransPatchGAN introduces contextual feature interaction into adversarial evaluation. A progressive ablation study compares Pix2Pix, NSNet-Base, NSNet-D, and TransNSNet. Image-level performance was summarized over a fixed convergence window (epochs 80--101) from one training run per model. TransNSNet achieved the best mean value in nine of ten metrics, whereas NSNet-Base achieved the highest visual saliency-induced index. For fixed-fraction high-intensity regions, NSNet-D was marginally higher in slice-level recall and balanced accuracy, but the equal-weight patient-level values were nearly identical and the highest Dice values shifted to TransNSNet. These differences were small relative to between-patient variability. In the separate adaptive patient-level analysis ($n=10$), no pairwise model difference remained statistically significant after Holm correction. TransNSNet also had a substantially smaller parameter footprint than the original next-scale models. These findings indicate promising image-quality trends, but they do not establish statistical superiority or clinical validity; synthesized PET-like images should not be interpreted as substitutes for clinically acquired PET.
\end{abstract}

\begin{keyword}
CT-to-PET image synthesis, medical image synthesis, multi-scale generation, Transformer, TransNSNet, TransPatchGAN.
\end{keyword}

\end{frontmatter}

\noindent\textit{Preprint. Submitted for publication.}\par\medskip

\section{Introduction}

Medical imaging plays a fundamental role in modern disease diagnosis, treatment planning, and longitudinal monitoring. Different imaging modalities provide complementary information about human anatomy and physiology. Computed tomography (CT) provides high-resolution anatomical and structural information by measuring tissue-dependent X-ray attenuation, making it suitable for depicting organs, bones, and tissue-density variations. Positron emission tomography (PET), in contrast, visualizes metabolic and functional activity through the spatial distribution of radiotracer uptake. In oncology, the combination of anatomical CT information and metabolic PET information is particularly valuable for tumor localization, staging, treatment response assessment, and follow-up analysis.

Despite its clinical value, PET imaging remains less accessible than CT in many practical scenarios. PET examinations require radiotracer administration, involve additional radiation exposure, depend on specialized imaging systems and radiochemistry infrastructure, and are generally more expensive than CT. As a result, CT examinations are more widely available, whereas paired PET information may be unavailable for many patients or difficult to obtain repeatedly. These limitations have motivated increasing interest in CT-to-PET image synthesis, which aims to estimate PET-like images from CT images and thereby provide auxiliary PET-like intensity representations for research and methodological analysis. It should be emphasized that synthesized PET-like images are not intended to replace clinically acquired PET scans. Instead, they may serve as complementary computational outputs when real PET acquisition is unavailable, limited, or impractical.

CT-to-PET synthesis is an intrinsically challenging cross-modality image translation problem. CT and PET are generated by fundamentally different physical mechanisms. CT mainly reflects anatomical density and attenuation properties, whereas PET reflects radiotracer uptake associated with biological and metabolic activity. Therefore, PET uptake patterns cannot be fully determined from CT anatomy alone. A reliable CT-to-PET synthesis model must learn statistical correspondences between the two modalities while avoiding physiologically implausible uptake patterns. This requirement is especially important in medical imaging, because synthetic images that appear visually plausible but contain incorrect metabolic information may mislead subsequent interpretation or downstream analysis.

Deep learning has substantially advanced medical image synthesis and cross-modality translation. Encoder-decoder networks and U-Net-based architectures can learn direct image-to-image mappings under paired supervision. However, regression-dominated training often produces overly smooth predictions and may suppress subtle local structures. Generative adversarial networks (GANs), especially conditional image-to-image translation frameworks such as Pix2Pix, introduce a discriminator to encourage generated images to better match the target-domain distribution. Nevertheless, conventional single-output translation frameworks still face difficulties in CT-to-PET synthesis. A single full-resolution supervision signal may be insufficient to guide the model to learn both global metabolic layout and local uptake details. In addition, the discriminator design affects what type of realism is emphasized during adversarial training. A purely local discriminator can provide regional feedback but may lack global contextual awareness, while a global discriminator may not sufficiently penalize local uptake inconsistencies.

Another practical challenge is the construction of paired CT-PET datasets. Supervised CT-to-PET translation requires well-matched source and target images. However, paired CT-PET data are not always directly available in a slice-level format suitable for training image translation models. In this study, we therefore construct a paired CT-PET translation dataset by screening, matching, and preprocessing CT and PET images from the public non-small cell lung cancer (NSCLC) Radiogenomics dataset \cite{bakr_data_nsclc_radiogenomics_2017,bakr_radiogenomic_dataset_2018}. Only cases with matched CT and PET slice correspondence are retained, enabling supervised evaluation of CT-to-PET synthesis models.

To address the above challenges, this study proposes TransNSNet, a Transformer-enhanced next-scale adversarial framework for CT-to-PET image synthesis. Pix2Pix is used as a representative conditional GAN baseline. NSNet-Base is constructed by replacing the Pix2Pix U-Net generator with a next-scale generator that produces PET predictions at multiple spatial scales and provides coarse-to-fine supervision during training. NSNet-D further replaces the PatchGAN-style discriminator with a Transformer-enhanced TransPatchGAN discriminator to isolate the contribution of discriminator-side contextual modeling. Finally, TransNSNet combines a Transformer-enhanced next-scale generator with the TransPatchGAN discriminator, enabling both generator-side global feature interaction and discriminator-side contextual adversarial feedback.

The main contributions of this study are summarized as follows:

\begin{enumerate}
\item We curate a paired CT-PET image translation dataset from the public NSCLC Radiogenomics collection. The contribution includes patient screening, slice-level CT-PET matching, image preprocessing, and patient-level train-test separation, resulting in 147 retained patients and 41,864 paired slices for supervised synthesis experiments.
\item We develop and analyze a Transformer-enhanced next-scale generator for CT-to-PET synthesis. The generator predicts PET-like images at six spatial resolutions and is optimized with coarse-to-fine reconstruction supervision; its contribution is evaluated separately for structural, perceptual, and error-based metrics rather than assumed to be uniform.
\item We introduce a Transformer Block into the discriminator to construct TransPatchGAN. This design extends conventional patch-level adversarial evaluation with contextual feature interaction and enables a component-wise analysis of local and broader uptake consistency.
\item We conduct a comprehensive evaluation including progressive ablation, fixed-fraction high-intensity-region and adaptive lesion-like uptake analyses, patient-level statistical testing, and model parameter comparison. TransNSNet obtains the best convergence-window mean in nine of ten image-quality metrics with a substantially smaller parameter footprint than the original next-scale models, while the regional results reveal model- and metric-dependent trade-offs.
\end{enumerate}

The remainder of this paper is organized as follows. Section 2 reviews related work on cross-modality medical image synthesis, GAN-based medical image translation, multi-scale generation, Transformer-based image modeling, and evaluation of synthetic medical images. Section 3 introduces the proposed framework and its ablation variants. Section 4 describes the dataset construction, preprocessing procedure, experimental settings, ablation design, and evaluation metrics. Section 5 presents quantitative and qualitative results. Section 6 discusses the technical advantages, limitations, and use boundaries of the proposed method. Section 7 concludes the paper and outlines future research directions.

\section{Related Work}

\begin{figure*}[ht]
\centering
\includegraphics[width=\textwidth,height=0.86\textheight,keepaspectratio]{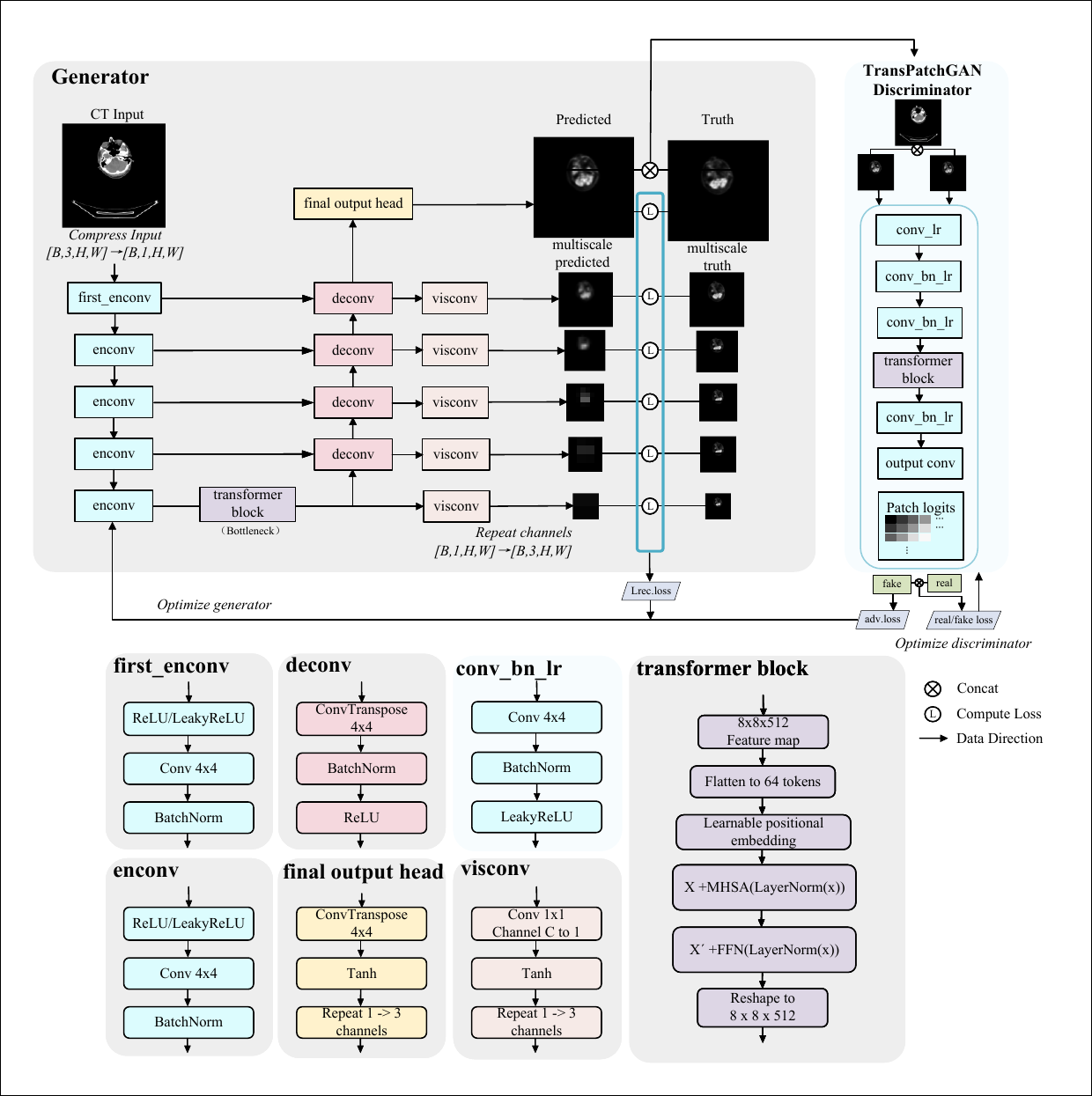}
\caption{Architecture of the proposed TransNSNet framework. The generator selects the middle CT channel, produces one full-resolution and five auxiliary PET predictions, and applies reconstruction supervision at six scales. TransPatchGAN evaluates real [CT, PET truth] and fake [CT, predicted PET] pairs in separate forward passes with shared weights.}
\label{fig:architecture}
\end{figure*}

\subsection{Generative medical image synthesis and modality translation}

Medical image synthesis has been widely studied as a way to mitigate data scarcity, reduce acquisition burden, and provide missing or enhanced imaging information. Early work has relied heavily on generative adversarial networks (GANs). PACGAN combines progressive growing with an auxiliary classifier GAN to generate high-resolution, class-conditional brain MR images for Alzheimer's disease and healthy-control classes \cite{lai_high-resolution_2025}. Thermogram-based breast cancer detection has also used a GAN module to synthesize critical regions of interest before downstream classification \cite{veerlapalli_hybrid_2025}. In lesion-level CT augmentation, StylePix2pix extends pix2pix with style-based one-to-many generation so that physician-drawn sketches can produce diverse lung tumor CT images \cite{toda_lung_2022}. These studies address limited data and class imbalance, but their primary objective is data augmentation or class-conditional synthesis rather than patient-specific cross-modality translation.

GAN-based synthesis has also been applied to clinically motivated image restoration and enhancement. Sparse-view CT artifact reduction has been formulated as a paired image restoration problem using conditional GANs, where CGANs improve CT-value restoration over AE and U-Net baselines but may still introduce hallucinated regions \cite{usui_reducing_2024}. Synthetic contrast-enhanced abdominal CT has been generated from non-enhanced CT through weakly supervised learning with virtual non-contrast CT, improving lesion detection sensitivity in radiologist evaluation \cite{han_clinical_2024}. For liver imaging, MAN-GAN uses mask-adaptive normalization and cycle-consistency to synthesize multiphase enhanced CT from non-contrast CT \cite{zhao_man-gan_2025}. More generally, dual-domain generative modeling has been proposed to couple acquisition-domain and image-domain constraints for reconstruction and synthesis tasks such as low-dose PET/CT reconstruction, metal artifact reduction, fast MRI reconstruction, and PET/CT synthesis \cite{zhang_generalized_2023}. U-Net-based restoration models such as U-Tunnel-Net further show the value of architecture-level design for preserving structure during denoising and reconstruction \cite{gokmen_effective_2025}. However, most of these methods are either intra-modality enhancement, organ-specific contrast synthesis, or reconstruction tasks. They do not directly address the anatomical-to-metabolic ambiguity inherent in CT-to-PET synthesis, where the target PET uptake pattern is not fully determined by CT anatomy.

\subsection{Diffusion models and cross-modality synthesis}

Diffusion models have recently become a strong alternative to GANs for medical image generation. Medfusion, a conditional latent denoising diffusion probabilistic model, was compared with StyleGAN-3, ProGAN, and cGAN across fundus, chest radiograph, and histopathology images, showing improved diversity while maintaining competitive fidelity \cite{muller-franzes_multimodal_2023}. Large-scale image-text generative foundation models such as MINIM further extend medical synthesis across organs and modalities, demonstrating that synthetic images can improve multiple downstream applications including diagnosis, report generation, and self-supervised learning \cite{wang_self-improving_2025}. At the same time, privacy analyses show that unconditional latent diffusion models can memorize patient imaging data, indicating that high-quality synthesis must be evaluated together with privacy and safety risks \cite{dar_unconditional_2025}.

For concrete cross-modality translation, Hu et al. proposed HUDiff for CBCT-to-CT synthesis, combining a ViT-enhanced U-Net diffusion backbone, information bottleneck compression, and dynamic skip modulation to improve CBCT image quality for radiotherapy \cite{hu_cbct--ct_2025}. Nguyen et al. introduced CPDM for CT-to-PET translation and constructed a large paired CT-PET dataset. CPDM integrates attention and attenuation maps as domain-knowledge conditions to guide the diffusion process toward metabolically relevant regions and PET correction cues \cite{nguyen_ct_2024}. These methods demonstrate the promise of diffusion-based cross-modality synthesis, but they also expose several practical constraints: diffusion models can be computationally expensive, their performance often depends on strong domain-specific conditional maps, and large paired datasets may not be available in many clinical settings. These constraints motivate efficient adversarial alternatives that can exploit paired supervision and multi-scale learning while remaining focused on the CT-to-PET task.

\subsection{Segmentation-oriented medical image modeling}

A second line of related work focuses on segmentation, where medical images are used to delineate lesions, organs, or anatomical regions. Although segmentation differs from image synthesis, recent segmentation studies demonstrate the importance of multi-scale context, attention, and modality-aware representation learning in medical imaging. Examples include CNN-Transformer hybrid segmentation with cross-layer fusion \cite{chen_dual_2025}, multi-modal brain tumor segmentation with multi-scale contextual aggregation and gated attention \cite{aslam_enhanced_2025}, conditional diffusion-based tumor segmentation \cite{yu_conditional_2025}, and modality-adaptive whole-body segmentation across imaging modalities \cite{chen_modality-projection_2025}. Other work addresses practical constraints such as limited annotation, acoustic artifacts, and privacy-preserving federated training \cite{enshaei_covid-rate_2022,zhang_satu-net_2025,skorupko_federated_2025,zhang_generative_2025}. Together, these studies provide architectural motivation for combining local detail modeling with broader contextual information. Nevertheless, segmentation objectives are mask-centric and do not require synthesis of a full target-domain intensity distribution. They therefore offer useful design insights but do not directly solve CT-to-PET image translation.

\subsection{Diagnostic classification and evaluation context}

Diagnostic classification studies provide another perspective on how synthesized or transformed images should ultimately be evaluated. Multi-task and attention-based classifiers show that image representations are often judged not only by visual appearance but also by their usefulness for disease classification, lesion localization, and interpretable decision support \cite{reddy_cxr-multitasknet_2025,ferdous_mednet_2025,slimi_trustworthy_2025}. Other studies combine synthetic data, Transformer-based global-context modeling, or multi-modal clinical information with diagnostic prediction, further emphasizing that generated or transformed images should be assessed according to task-relevant utility rather than visual similarity alone \cite{brahmareddy_transbreastnet_2025,s_web_2025,parathanath_cbmnet_2025,sujini_automated_2025}. Beyond image classifiers, clinical reasoning benchmarks and cohort-level prognostic studies show that final accuracy or immediate visual appearance may be insufficient without evaluating reasoning completeness, treatment-planning relevance, or longitudinal outcome associations \cite{qiu_quantifying_2025,zhu_prognostic_2025}. For CT-to-PET synthesis, these observations imply that image-level metrics such as SSIM and PSNR should be interpreted cautiously unless synthetic PET-like images also preserve lesion-relevant uptake patterns and downstream task utility.

\subsection{Explainability, attribution, privacy, and reliability}

As medical generative models are explored for decision-support research, reliability and interpretability become critical. Visual attribution and generative counterfactual auditing can help identify diagnostically relevant image regions or undesirable classifier shortcuts, but these methods depend on the faithfulness of the generated images themselves \cite{siddiqui_latent_2025,degrave_auditing_2023}. If a generative model changes lesion-relevant content while preserving visual plausibility, attribution or counterfactual explanations may become misleading. Privacy and evaluation risks are also important: memorization analyses show that synthetic medical images may reproduce patient data rather than generate novel samples \cite{dar_unconditional_2025}, and broader clinical AI evaluations show that apparently accurate outputs can still miss important reasoning or outcome-related evidence \cite{qiu_quantifying_2025,zhu_prognostic_2025}. These observations motivate a conservative interpretation of synthetic PET-like images: a visually plausible CT-to-PET output should be regarded as a technical output whose lesion-level correctness and downstream utility require additional evaluation.

\subsection{Summary and motivation}

The above studies reveal four limitations that directly motivate this work. First, GAN-based medical synthesis methods are effective for augmentation, enhancement, and organ-specific image generation, but many of them are designed for intra-modality or task-specific settings rather than CT-to-PET translation \cite{lai_high-resolution_2025,veerlapalli_hybrid_2025,toda_lung_2022,zhao_man-gan_2025}. Second, diffusion models provide strong fidelity and diversity for cross-modality synthesis, including CBCT-to-CT and CT-to-PET, but they often require heavy computation, large paired datasets, and carefully designed domain-specific conditions \cite{muller-franzes_multimodal_2023,hu_cbct--ct_2025,nguyen_ct_2024}. Third, segmentation and classification studies demonstrate the value of multi-scale context, attention, modality adaptation, and downstream task coupling, yet their objectives do not directly optimize full PET-like image synthesis \cite{chen_dual_2025,aslam_enhanced_2025,chen_modality-projection_2025,reddy_cxr-multitasknet_2025}. Fourth, explainability and privacy studies show that synthetic medical images must be evaluated beyond visual similarity because generative models may hallucinate, memorize, or support misleading interpretations \cite{siddiqui_latent_2025,degrave_auditing_2023,dar_unconditional_2025}.

In response, this study focuses on CT-to-PET image synthesis through TransNSNet, a Transformer-enhanced next-scale adversarial framework. The next-scale generator provides explicit coarse-to-fine supervision so that global uptake layout and local intensity structure can be learned at multiple spatial levels. The TransPatchGAN discriminator introduces contextual adversarial feedback, aiming to compensate for the limited global awareness of purely local PatchGAN-style discrimination. By comparing Pix2Pix, NSNet-Base, NSNet-D, and TransNSNet under a progressive ablation protocol, this work evaluates how next-scale generation and Transformer-enhanced adversarial modeling contribute to PET-like image synthesis from CT.

\section{Method}

\subsection{Problem Formulation}

Let $\mathcal{X}$ denote the source CT domain and $\mathcal{Y}$ denote the target PET domain. For each paired training sample $(x_i,y_i)$, $x_i \in \mathcal{X}$ represents a CT image and $y_i \in \mathcal{Y}$ represents the corresponding PET image. In this study, both CT and PET images are represented as three-channel images with a spatial resolution of 256 $\times$ 256. The goal of CT-to-PET synthesis is to learn a generator $G: \mathcal{X} \rightarrow \mathcal{Y}$ that estimates a PET-like image
\begin{equation}
\hat{y}_i = G(x_i)
\end{equation}
from the CT input $x_i$.

Unlike natural image translation, CT-to-PET synthesis is a cross-modality mapping between anatomical and metabolic imaging. CT mainly reflects tissue attenuation and anatomical structure, whereas PET reflects radiotracer uptake and metabolic activity. Therefore, the mapping from CT to PET is not a simple appearance conversion. A suitable model should preserve anatomical correspondence with the CT input while approximating the global intensity distribution and local uptake patterns of real PET images.

The supervised training objective is formulated using paired CT-PET images. The generated PET image is compared with the real PET image at multiple spatial scales, and adversarial learning is used to encourage the generated PET image to follow the visual and distributional characteristics of real PET images.

\subsection{Overview of the Progressive Framework}

The experimental framework consists of four progressively constructed models: Pix2Pix, NSNet-Base, NSNet-D, and TransNSNet.

First, Pix2Pix is used as the baseline conditional GAN. It uses a U-Net generator and a PatchGAN discriminator for paired image-to-image translation. Pix2Pix provides a standard reference for describing metric differences associated with the proposed next-scale generator and Transformer-enhanced components.

Second, NSNet-Base replaces the Pix2Pix U-Net generator with a next-scale generator, denoted as $G_{\mathrm{next}}$. In this work, next-scale generation refers to a specific multi-scale prediction and supervision strategy: the generator produces one full-resolution PET prediction and auxiliary PET predictions at progressively lower spatial resolutions, and each output is supervised by a PET target at the same scale. Thus, ``multi-scale'' describes the use of several spatial resolutions, whereas ``next-scale'' denotes the concrete generator design evaluated in this study. The discriminator used in NSNet-Base follows a PatchGAN-style conditional discriminator structure. The comparison between Pix2Pix and NSNet-Base provides a component-wise assessment of the next-scale generator.

Third, NSNet-D keeps the original next-scale generator $G_{\mathrm{next}}$ and replaces the PatchGAN-style discriminator with the Transformer-enhanced TransPatchGAN discriminator. This model provides a component-wise assessment of discriminator-side Transformer enhancement while keeping the generator unchanged.

Finally, TransNSNet replaces the original next-scale generator with a Transformer-enhanced generator, denoted as $G_{\mathrm{TransNSNet}}$, while retaining the TransPatchGAN discriminator. This final model introduces Transformer-based contextual modeling into both the generator-side feature representation and the adversarial feedback process. TransNSNet is therefore regarded as the proposed final framework in this study.

Figure~\ref{fig:architecture} summarizes the final TransNSNet architecture. The generator produces PET-like predictions at multiple scales, while the full-resolution output is evaluated by the conditional TransPatchGAN discriminator together with the CT input.

\subsection{NSNet-Base: Next-Scale Generator}

The generator of NSNet-Base, denoted as $G_{\mathrm{next}}$, follows an encoder-decoder design with multiple prediction outputs at different spatial scales. The encoder extracts hierarchical CT features by progressively reducing spatial resolution and increasing feature abstraction. The decoder reconstructs PET-like representations from these encoded features.

Instead of producing only a single final PET image, the generator produces PET predictions at multiple scales. In the current implementation, the supervised output scales correspond to 256 $\times$ 256, 128 $\times$ 128, 64 $\times$ 64, 32 $\times$ 32, 16 $\times$ 16, and 8 $\times$ 8. The real PET image is downsampled to each corresponding scale and used as the target for the generated prediction at that scale.

This next-scale design provides coarse-to-fine supervision. Low-resolution predictions are expected to capture the broad metabolic uptake distribution and global image layout, while high-resolution predictions refine local intensity transitions and regional details. In this way, the generator receives direct learning signals from multiple spatial levels rather than relying only on the final full-resolution output.

\subsection{TransPatchGAN Discriminator}

The PatchGAN-style conditional discriminator receives the concatenation of CT and PET images along the channel dimension. For real samples, the input is the CT image concatenated with the real PET image; for generated samples, the input is the CT image concatenated with the generated PET image. This conditional design allows the discriminator to evaluate whether the generated PET image is realistic under the anatomical condition provided by the CT input.

To extend the adversarial feedback design, we introduce a Transformer-enhanced TransPatchGAN discriminator. TransPatchGAN retains the patch-level discrimination principle of PatchGAN but introduces Transformer-based contextual modeling into the discriminator. In the implemented discriminator, the Transformer Block is inserted after the convolutional feature-extraction layers and before the final patch-logit prediction layer. The convolutional feature map is treated as a sequence of spatial tokens for self-attention, and the resulting contextual representation is then mapped to patch-level logits. This design aims to provide both local patch-level realism assessment and broader contextual awareness, which is evaluated empirically rather than assumed from the architecture.

In NSNet-D, TransPatchGAN is introduced while the generator remains $G_{\mathrm{next}}$. Therefore, NSNet-D isolates the effect of discriminator enhancement. In TransNSNet, TransPatchGAN is used together with the Transformer-enhanced generator, forming the final model.

\subsection{TransNSNet: Transformer-Enhanced Next-Scale Framework}

TransNSNet is the final proposed model in this study. It combines a Transformer-enhanced next-scale generator with the TransPatchGAN discriminator. The motivation is that CT-to-PET synthesis requires not only local texture reconstruction but also global contextual understanding. PET uptake patterns are spatially distributed and may depend on broader anatomical and physiological contexts. Therefore, adding Transformer-based modeling to the generator may help the model capture long-range dependencies beyond local convolutional receptive fields.

The Transformer-enhanced generator builds upon the next-scale generator structure. It preserves the multi-scale PET prediction mechanism and introduces a Transformer Block at the generator bottleneck, where the spatial representation is most compact after encoder downsampling and before decoder reconstruction. This placement enhances global feature interaction before the decoder produces the full-resolution and auxiliary PET predictions. The generated multi-scale PET predictions are supervised by downsampled PET targets, as in NSNet-Base. The final full-resolution PET prediction is further evaluated by the TransPatchGAN discriminator.

Compared with NSNet-Base, TransNSNet modifies both sides of the adversarial framework. The generator is enhanced with Transformer-based contextual modeling, and the discriminator is replaced with TransPatchGAN. Compared with NSNet-D, TransNSNet assesses metric changes associated with adding Transformer enhancement to the generator when the discriminator has already been enhanced. Therefore, TransNSNet should be interpreted as an integrated Transformer-enhanced next-scale adversarial framework rather than a simple combination of NSNet-Base and Transformer modules.

\subsection{Training Objective}

All NSNet-based models are trained using a unified protocol derived from the original next-scale implementation. The objective combines adversarial learning at full resolution with reconstruction supervision at multiple scales. Each training sample consists of a normalized three-channel CT input, denoted by $x$, and its paired ground-truth PET image, denoted by $y$. Both images have a spatial resolution of $256\times256$ pixels and intensity values in the range $[0,1]$.

The notation for multi-scale prediction and supervision is summarized as
\begin{equation}
\begin{aligned}
\mathcal{S} &= \{256,128,64,32,16,8\}, \\
\hat{y}_s &= G_s(x), \qquad s\in\mathcal{S}, \\
y_s &= R_s(y), \qquad s\in\mathcal{S}.
\end{aligned}
\label{eq:multiscale-notation}
\end{equation}
Here, $\mathcal{S}$ is the set of spatial output sizes, and $s$ identifies an $s\times s$ resolution. The function $G_s$ denotes the generator output branch at scale $s$, and $\hat{y}_s$ is its PET prediction. The resizing operator $R_s$ produces the scale-matched supervision target $y_s$ from the ground-truth PET image. At full resolution, $\hat{y}_{256}$ is the final generator output and $y_{256}=y$.

For two tensors $a$ and $b$ with identical dimensions, 
$\mathrm{MSE}(a,b)$ denotes the mean squared error between their 
corresponding elements, $\mathcal{L}_{1}(a,b)$ denotes the mean 
absolute error used for pixel-wise reconstruction, and 
$\mathcal{L}_{2}(a,b)$ denotes the mean squared error used for 
pixel-wise reconstruction. These losses are defined as
\begin{align}
\mathrm{MSE}(a,b)
&=
\frac{1}{N}
\sum_{i=1}^{N}
(a_i-b_i)^2, \\
\mathcal{L}_{1}(a,b)
&=
\frac{1}{N}
\sum_{i=1}^{N}
|a_i-b_i|, \\
\mathcal{L}_{2}(a,b)
&=
\frac{1}{N}
\sum_{i=1}^{N}
(a_i-b_i)^2,
\end{align}
where $N$ is the total number of scalar elements in each tensor,
$i$ is the element index, and $a_i$ and $b_i$ denote the $i$th
elements of $a$ and $b$, respectively. In the adversarial objective,
$a$ represents the discriminator score map and $b$ represents its
corresponding real or fake target-label tensor. In the reconstruction
objective, $a$ represents a generated PET image and $b$ represents
the scale-matched ground-truth PET image. Although
$\mathrm{MSE}(a,b)$ and $\mathcal{L}_{2}(a,b)$ have the same
mathematical form, the two notations distinguish their roles:
$\mathrm{MSE}$ is applied to discriminator outputs for adversarial
learning, whereas $\mathcal{L}_{2}$ is applied to image intensities
for PET reconstruction.

At full resolution, the conditional discriminator $D$ receives either the real pair $(x,y)$ or the generated pair $(x,\hat{y}_{256})$ and produces a patch-wise score map. Let $\mathbf{1}$ and $\mathbf{0}$ denote tensors of ones and zeros, respectively, with the same dimensions as that score map. Following the least-squares generative adversarial network formulation, the discriminator loss $\mathcal{L}_{D}$ is
\begin{equation}
\mathcal{L}_{D}
= \frac{1}{2}
\left[
\mathrm{MSE}\!\left(D(x,y),\mathbf{1}\right)
+
\mathrm{MSE}\!\left(D(x,\hat{y}_{256}),\mathbf{0}\right)
\right].
\end{equation}

Here, $\mathcal{L}_{D}$ encourages $D$ to assign real and generated CT--PET pairs to the target labels $\mathbf{1}$ and $\mathbf{0}$, respectively. The generator adversarial loss $\mathcal{L}_{\mathrm{adv}}$ is
\begin{equation}
\mathcal{L}_{\mathrm{adv}}
=
\mathrm{MSE}\!\left(D(x,\hat{y}_{256}),\mathbf{1}\right),
\end{equation}
where $\mathcal{L}_{\mathrm{adv}}$ encourages the full-resolution generated pair to be classified as real. Combining adversarial and reconstruction supervision, the full-resolution generator loss $\mathcal{L}_{G}^{256}$ is
\begin{equation}
\mathcal{L}_{G}^{256}
=
\mathcal{L}_{\mathrm{adv}}
+
\frac{1}{2}
\left[
\mathcal{L}_{1}(\hat{y}_{256},y)
+
\mathcal{L}_{2}(\hat{y}_{256},y)
\right].
\end{equation}

At lower resolutions, no adversarial term is applied. Let $\mathcal{S}_{\mathrm{aux}}=\mathcal{S}\setminus\{256\}=\{128,64,32,16,8\}$ denote the auxiliary scale set. For each $s\in\mathcal{S}_{\mathrm{aux}}$, the auxiliary reconstruction loss $\mathcal{L}_{G}^{s}$ is
\begin{equation}
\mathcal{L}_{G}^{s}
=
\frac{1}{2}
\left[
\mathcal{L}_{1}(\hat{y}_{s},y_{s})
+
\mathcal{L}_{2}(\hat{y}_{s},y_{s})
\right].
\end{equation}
The complete generator objective $\mathcal{L}_{G}$ is
\begin{equation}
\mathcal{L}_{G}
=
\mathcal{L}_{G}^{256}
+
\sum_{s \in \mathcal{S}_{\mathrm{aux}}}
\mathcal{L}_{G}^{s}.
\end{equation}
Here, $\mathcal{L}_{G}^{s}$ denotes the reconstruction loss at scale $s$, and $\mathcal{L}_{G}$ is the scalar objective minimized when updating the generator. The factor $1/2$ assigns equal averaging weights to the $L_1$ and $L_2$ reconstruction terms, and the auxiliary-scale losses are summed with equal weights. This loss protocol is held fixed across NSNet-Base, NSNet-D, and TransNSNet; consequently, the ablation study isolates architectural changes rather than changes in the training objective.

\subsection{Implementation Details}

All models are implemented in PyTorch. The input CT images and target PET images are represented as three-channel images with a spatial size of 256 $\times$ 256. Image intensities are normalized to the range [0, 1] before training. The models are trained from scratch without preloaded generator or discriminator checkpoints.

All compared models are trained with a batch size of 1. Adam optimization is used with an initial learning rate of 0.0002 and momentum parameters $\beta_1 = 0.5$ and $\beta_2 = 0.999$. In each evaluation log, epoch 0 denotes the pre-training evaluation; each model is subsequently trained for 101 epochs (epochs 1--101). Checkpoint-monitoring evaluations are performed separately on the training split and the patient-level test split listed in Table~\ref{tab:dataset}, and model checkpoints are saved during training.

For NSNet-Base, NSNet-D, and TransNSNet, the multi-scale generation process follows the scale order 256 $\times$ 256, 128 $\times$ 128, 64 $\times$ 64, 32 $\times$ 32, 16 $\times$ 16, and 8 $\times$ 8. The same dataset, preprocessing protocol, optimizer settings, and training objective are used across models whenever possible to ensure fair comparison.

Each model was represented by one training run; repeated training with different random seeds was not performed. Parts of the experiments were run on different computers, but the same paired dataset, preprocessing procedure, optimization settings, and evaluation conditions were used. NSNet-Base training was interrupted once by a system restart after the epoch-87 evaluation. Training was resumed from the saved weights and completed through epoch 101. Its pre- and post-interruption evaluation CSV files were manually concatenated using continuous epoch numbers without altering evaluation values. Pix2Pix, NSNet-D, and TransNSNet completed without training interruption, and their evaluation CSV files were not manually merged.

\section{Experiments}

\subsection{Dataset}

\begin{figure*}[h]
\centering
\includegraphics[width=\textwidth,height=0.86\textheight,keepaspectratio]{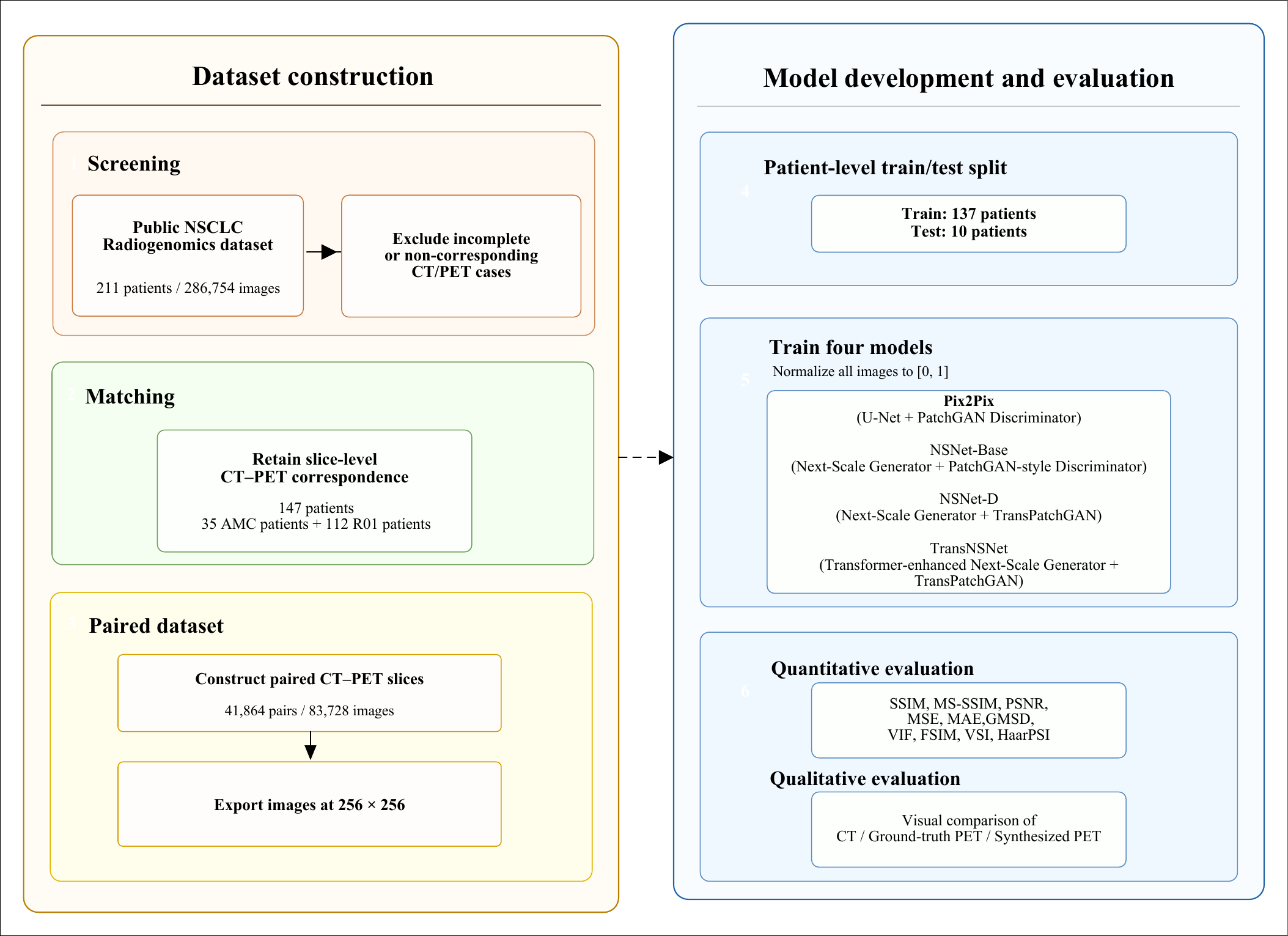}
\caption{Study workflow for paired CT-PET dataset construction and experimental splitting.}
\label{fig:study-workflow}
\end{figure*}

The experiments were conducted using CT and PET images derived from the public NSCLC Radiogenomics dataset \cite{bakr_data_nsclc_radiogenomics_2017,bakr_radiogenomic_dataset_2018}. Although the original collection contains multimodal imaging data, not all cases are directly suitable for supervised CT-to-PET image translation. Since the proposed models require paired source and target images, patients whose CT and PET images could not be matched in a slice-level one-to-one manner were excluded. The overall construction workflow is summarized in Figure~\ref{fig:study-workflow}.

After data screening, 147 patients were retained for this study. The retained cohort contained two subsets, AMC and R01. For each included patient, the number of CT slices was identical to the number of PET slices, enabling direct slice-level pairing between the two modalities. The final dataset contained 41,864 paired CT-PET slices, corresponding to 83,728 images in total. The dataset was divided at the patient level into training and test subsets to avoid overlap of slices from the same patient across different subsets. Table~\ref{tab:dataset} provides the detailed split-level statistics.

\begin{table*}[h]
\centering
\caption{Constructed paired CT-PET dataset after screening and patient-level splitting.}
\label{tab:dataset}
\scriptsize
\setlength{\tabcolsep}{2pt}
\renewcommand{\arraystretch}{1.1}

\begin{tabularx}{\textwidth}{l*{6}{>{\raggedleft\arraybackslash}X}}
\toprule
& \multicolumn{2}{c}{AMC}
& \multicolumn{2}{c}{R01}
& \multicolumn{2}{c}{Total} \\
\cmidrule(lr){2-3}
\cmidrule(lr){4-5}
\cmidrule(lr){6-7}
Split
& Patients
& CT-PET pairs
& Patients
& CT-PET pairs
& Patients
& Images \\
\midrule
Train & 32 & 8,590 & 105 & 30,369 & 137 & 77,918 \\
Test  & 3  & 765   & 7   & 2,140  & 10  & 5,810  \\
Total & 35 & 9,355 & 112 & 32,509 & 147 & 83,728 \\
\bottomrule
\end{tabularx}
\end{table*}

The dataset construction procedure is an important part of this study because supervised CT-to-PET synthesis depends on reliable paired samples. Inaccurate pairing between CT and PET slices would introduce noise into the learning target and weaken the reliability of quantitative evaluation.

\subsection{Data Screening and Preprocessing}

The data screening procedure was designed to construct reliable paired samples for supervised cross-modality translation. First, patients with incomplete or non-corresponding CT-PET image sequences were removed. Second, for the remaining patients, CT and PET slices were matched according to the same patient and the same slice order. Because the retained cases had identical CT and PET slice counts, this procedure produced paired training samples in the form of $(x,y)$, where $x$ denotes the CT image and $y$ denotes the corresponding PET image.

All images used in the experiments had a spatial size of 256 $\times$ 256 pixels after export. Therefore, no resizing operation was applied during preprocessing. CT images were exported as three-channel image files using MicroDicom and were directly used as the source input of the network. In this workflow, the three channels are image-format channels rather than three independent anatomical sequences. The retained slice pairs rely on patient-level and slice-order correspondence, and no additional deformable registration was performed in the current preprocessing pipeline. Before model training, both CT and PET images were normalized to the intensity range [0, 1]. This normalization ensured that reconstruction losses were computed on a consistent numerical scale across modalities. Because the workflow uses normalized image exports rather than raw SUV-calibrated PET volumes, the synthesized outputs should be interpreted as normalized PET-like intensity maps rather than quantitative SUV estimates.

The proposed task was formulated as paired CT-to-PET synthesis. During training, the CT image served as the source-domain input, while the paired PET image served as the target-domain supervision. For NSNet-based models, the target PET image was additionally downsampled to multiple spatial resolutions to provide multi-scale supervision for the generator.

\subsection{Compared Models and Ablation Design}

To evaluate the proposed framework, we designed a progressive ablation study with four models: Pix2Pix, NSNet-Base, NSNet-D, and TransNSNet. This design allows component-wise metric comparisons at each architectural step.

Pix2Pix was used as the baseline model. It consists of a U-Net generator and a PatchGAN discriminator and represents a standard conditional GAN framework for paired image-to-image translation. This model provides a reference for describing metric changes associated with the proposed next-scale generator.

NSNet-Base replaces the Pix2Pix U-Net generator with $G_{\mathrm{next}}$, the next-scale generator. The discriminator follows a PatchGAN-style conditional discriminator structure. The comparison between Pix2Pix and NSNet-Base is used to describe how multi-scale PET prediction and next-scale supervision affect individual CT-to-PET synthesis metrics.

NSNet-D keeps the original next-scale generator unchanged and replaces the PatchGAN-style discriminator with TransPatchGAN. This experiment provides a component-wise comparison under the same generator and training protocol. Because each model was trained only once, observed differences are interpreted as performance trends associated with the discriminator change rather than as statistically established causal effects.

TransNSNet is the final proposed model. It combines the Transformer-enhanced next-scale generator with the TransPatchGAN discriminator. Compared with NSNet-D, TransNSNet further modifies the generator by introducing Transformer-based contextual modeling. This comparison describes the metric changes observed after generator-side Transformer enhancement when the discriminator has already been enhanced.

The ablation logic is summarized as follows: Pix2Pix to NSNet-Base compares the next-scale generator with the U-Net baseline; NSNet-Base to NSNet-D compares the Transformer-enhanced discriminator with the PatchGAN-style discriminator; NSNet-D to TransNSNet compares the Transformer-enhanced generator with the original next-scale generator; and Pix2Pix to TransNSNet compares the complete proposed framework with the baseline.

\begin{figure*}[h]
\centering
\includegraphics[width=\textwidth,height=0.30\textheight,keepaspectratio]{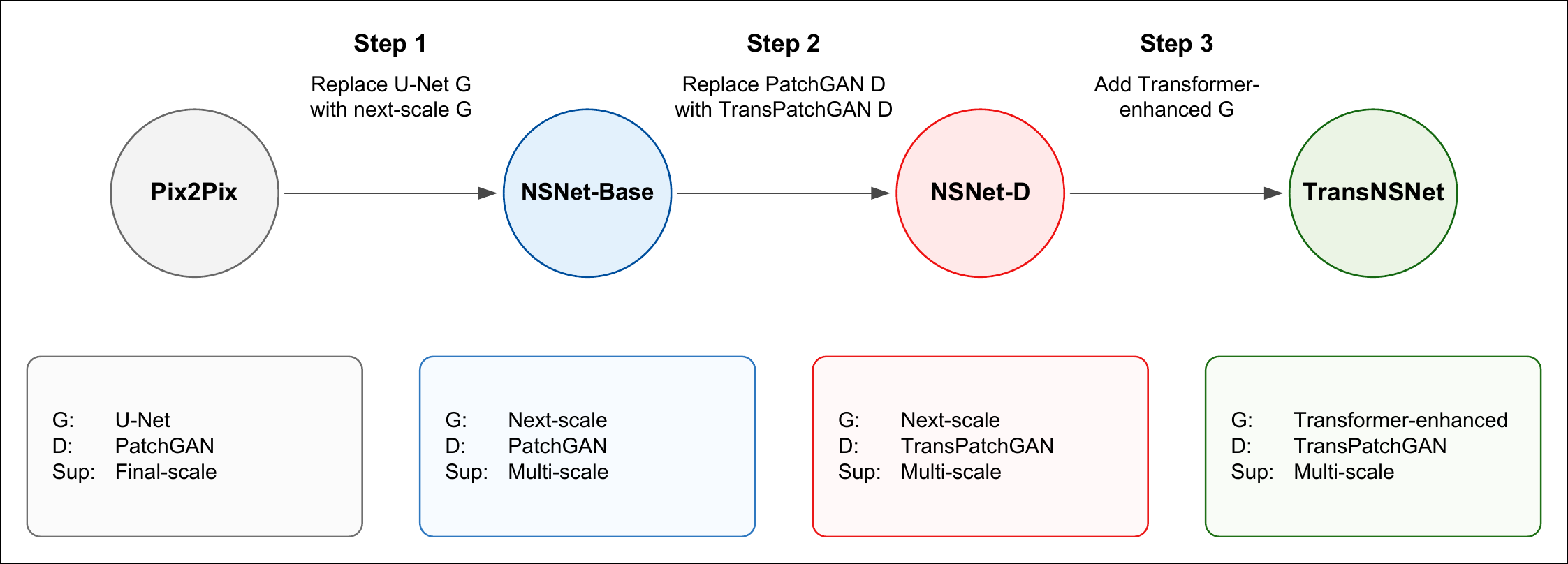}
\caption{Progressive ablation design from the Pix2Pix baseline to the final TransNSNet framework.}
\label{fig:ablation-design}
\end{figure*}

This progressive design helps avoid attributing performance changes to multiple architectural modifications at once. Table~\ref{tab:ablation-definitions} lists the model definitions used in the ablation study, and Figure~\ref{fig:ablation-design} illustrates the stepwise component changes.

\begin{table*}[htb]
\centering
\caption{Model definitions for the progressive ablation study.}
\label{tab:ablation-definitions}
\scriptsize
\setlength{\tabcolsep}{2pt}
\renewcommand{\arraystretch}{1.15}
\begin{tabularx}{\textwidth}{p{0.10\textwidth}p{0.16\textwidth}p{0.15\textwidth}>{\centering\arraybackslash}p{0.09\textwidth}>{\centering\arraybackslash}p{0.10\textwidth}>{\centering\arraybackslash}p{0.11\textwidth}Y}
\toprule
Model & Generator & Discriminator & \shortstack{Multi-scale\\supervision} & \shortstack{Transformer\\in generator} & \shortstack{Transformer\\in discriminator} & Ablation purpose \\
\midrule
Pix2Pix & U-Net generator & PatchGAN discriminator & -- & -- & -- & Baseline conditional GAN for paired CT-to-PET translation \\
NSNet-Base & Next-scale generator & PatchGAN-style discriminator & $\checkmark$ & -- & -- & Compare next-scale generation and multi-scale PET supervision with the baseline \\
NSNet-D & Next-scale generator & TransPatchGAN & $\checkmark$ & -- & $\checkmark$ & Compare TransPatchGAN with the PatchGAN-style discriminator \\
TransNSNet & Transformer-enhanced next-scale generator & TransPatchGAN & $\checkmark$ & $\checkmark$ & $\checkmark$ & Final proposed model; evaluate the complete Transformer-enhanced next-scale adversarial framework \\
\bottomrule
\end{tabularx}
\renewcommand{\arraystretch}{1.0}
\end{table*}

\subsection{Evaluation Metrics}

The synthesized PET images were evaluated using multiple image-quality metrics to assess reconstruction accuracy, structural similarity, perceptual quality, and local visual consistency. Specifically, we used structural similarity index (SSIM), multi-scale structural similarity index (MS-SSIM), peak signal-to-noise ratio (PSNR), mean squared error (MSE), mean absolute error (MAE), gradient magnitude similarity deviation (GMSD), visual information fidelity (VIF), feature similarity index measure (FSIM), visual saliency-induced index (VSI), and Haar wavelet-based perceptual similarity index (HaarPSI).

For SSIM, MS-SSIM, PSNR, VIF, FSIM, VSI, and HaarPSI, higher values indicate better synthesis quality. For MSE, MAE, and GMSD, lower values indicate better agreement between generated PET images and ground-truth PET images. Using this combination of metrics allows the evaluation to cover both pixel-wise reconstruction fidelity and perceptual/structural similarity.

A fixed-fraction high-intensity uptake (HU) definition was additionally used to assess agreement between PET-derived and synthesized percentile masks. Hereafter, the prefix HU denotes metrics computed under this fixed-fraction definition. For a PET image $y$ and a synthesized image $\hat{y}$, the masks were obtained by percentile thresholding each normalized slice. Let $M_y^{(q)}$ and $M_{\hat{y}}^{(q)}$ denote the binary masks corresponding to the top $q\%$ intensity pixels of $y$ and $\hat{y}$, respectively. The main fixed-fraction analysis used $q=5$, while threshold-sensitive Dice scores were also reported for $q=3$ and $q=10$.

After percentile thresholding within the ground-truth PET-derived body mask, connected components were identified independently in each 2D slice using 4-connectivity, and all components containing at least 20 pixels were retained for both the ground-truth and predicted masks; if filtering would otherwise yield an empty mask, the original thresholded mask was retained.

For the top-5\% masks, overlap quality was measured by Dice, IoU, precision, recall, specificity, and balanced accuracy. Dice and IoU were defined as
\begin{equation}
\mathrm{Dice}=\frac{2|M_y^{(5)} \cap M_{\hat{y}}^{(5)}|}{|M_y^{(5)}|+|M_{\hat{y}}^{(5)}|},
\end{equation}
\begin{equation}
\mathrm{IoU}=\frac{|M_y^{(5)} \cap M_{\hat{y}}^{(5)}|}{|M_y^{(5)} \cup M_{\hat{y}}^{(5)}|}.
\end{equation}
The intensity-error metrics denoted HU-MAE, HU-RMSE, and HU-peak error were computed inside the ground-truth top-5\% mask $M_y^{(5)}$. Centroid distance, peak-location distance, HD95, and ASSD were used to evaluate spatial and boundary consistency. False uptake area ratio and false uptake burden quantified predicted top-5\% regions outside the ground-truth top-5\% mask.

Two aggregation schemes were used for this fixed-fraction definition. Table~\ref{tab:high-uptake} reports slice-level means and standard deviations from all 2,905 test slices. Supplementary Table~\ref{tab:supp-top5-patient} reports the same 17 metrics after first averaging slices within each patient and then averaging the ten patient-level values with equal patient weights; its variability is the between-patient standard deviation.

A complementary patient-level analysis used an adaptive lesion-like uptake (LU) definition rather than a fixed pixel fraction. For each normalized 2D slice, the ground-truth and predicted images were independently thresholded at 50\% of their respective slice maxima; all 4-connected components containing at least 20 pixels were retained rather than only the largest component, no empty-mask fallback was applied, and the slice was included only if the filtered ground-truth mask contained at least 30 pixels. This yielded 2,744 evaluable slices and excluded 161 of 2,905 slices (5.5\%). Within each of the ten patients, metrics were averaged across evaluable slices, after which patients were assigned equal weight. The paired statistical analysis comprised the five LU metrics reported in Table~\ref{tab:adaptive-patient}: LU-Dice, LU-IoU, LU-Precision, LU-Recall, and LU-MAE. For each metric, TransNSNet was compared with each of the other three models using a two-sided Wilcoxon signed-rank test at the patient level, yielding 15 paired comparisons in total; Holm correction was applied across all 15 comparisons. Because this adaptive definition differs from the top-5\% fixed-fraction definition, values from the two analyses should not be compared directly.

In addition to quantitative metrics, qualitative visual comparisons were performed. The visual comparison includes CT input images, real PET images, and synthesized PET images from each model. These comparisons provide a qualitative assessment of overall PET-like intensity appearance and local high-intensity patterns.

\subsection{Experimental Settings}

All compared models were trained and evaluated using the same paired CT-PET dataset and preprocessing protocol. The input image size was 256 $\times$ 256 $\times$ 3. CT images were used as the source-domain input, and PET images were used as the target-domain output. All images were normalized to [0, 1]. All compared models were trained with a batch size of 1 using Adam with an initial learning rate of 0.0002, $\beta_1 = 0.5$, and $\beta_2 = 0.999$. In each evaluation log, epoch 0 denotes the pre-training evaluation; each model was subsequently trained for 101 epochs (epochs 1--101). For NSNet-Base, NSNet-D, and TransNSNet, adversarial MSE was applied at full resolution and reconstruction supervision using 0.5 $\times$ (L1 + L2) was applied at multiple scales.

The primary image-level comparison in Table~\ref{tab:main-quantitative} uses a fixed convergence window comprising epochs 80--101 (22 evaluations) for every model. Each entry is the mean over these epochs, and the accompanying standard deviation quantifies epoch-to-epoch fluctuation within that single training run. It is neither a between-patient or between-seed standard deviation nor a confidence interval. The fixed window avoids selecting an individual epoch according to test-set performance and can be specified without consulting which monitored test epoch gives the highest metric value.

In contrast, Table~\ref{tab:high-uptake} and Figures~\ref{fig:high-uptake-contour}--\ref{fig:error-maps} were generated from saved, model-specific checkpoints retained under the original best-SSIM monitoring rule: epoch 51 for Pix2Pix, epoch 12 for NSNet-Base, epoch 18 for NSNet-D, and epoch 67 for TransNSNet. Thus, the NSNet-Base regional and qualitative results use epoch 12. These checkpoint-based results have a different model-state selection and aggregation scheme from the convergence-window summary in Table~\ref{tab:main-quantitative} and are reported as complementary descriptive analyses.

\section{Results}

\subsection{Main Quantitative Comparison}

The main quantitative comparison summarizes the fixed convergence window from epochs 80--101. Table~\ref{tab:main-quantitative} reports the mean and epoch-to-epoch standard deviation over these 22 evaluations from the single training run of each model. The variability term describes within-run temporal fluctuation only and is not a confidence interval or an estimate of patient- or seed-level uncertainty.

TransNSNet achieved the best mean value in nine of the ten image-level metrics. NSNet-Base achieved the highest VSI mean (0.9346), compared with 0.9342 for TransNSNet. Accordingly, the results indicate a broad image-quality trend in favor of TransNSNet, but not uniform superiority across every metric.

\begin{table*}[hbt]
\centering
\caption{Main image-level comparison over the fixed convergence window (epochs 80--101; 22 evaluations per model). Values are mean $\pm$ epoch-to-epoch standard deviation from one training run. The standard deviation is not a between-patient or between-seed estimate and is not a confidence interval. The best mean value in each row is shown in bold.}
\label{tab:main-quantitative}
\scriptsize
\setlength{\tabcolsep}{2pt}
\renewcommand{\tabularxcolumn}[1]{m{#1}}
\begin{tabularx}{\textwidth}{l*{4}{>{\raggedleft\arraybackslash}X}}
\toprule
Metric & Pix2Pix
       & NSNet-Base
       & NSNet-D
       & TransNSNet \\
\midrule
SSIM $\uparrow$    & 0.8888 $\pm$ 0.0011 & 0.8883 $\pm$ 0.0010 & 0.8894 $\pm$ 0.0006 & \textbf{0.8918 $\pm$ 0.0015} \\
MS-SSIM $\uparrow$ & 0.8584 $\pm$ 0.0014 & 0.8588 $\pm$ 0.0002 & 0.8593 $\pm$ 0.0004 & \textbf{0.8649 $\pm$ 0.0011} \\
PSNR $\uparrow$    & 26.6841 $\pm$ 0.0339 & 26.8227 $\pm$ 0.0184 & 26.8493 $\pm$ 0.0156 & \textbf{27.1390 $\pm$ 0.0994} \\
MSE $\downarrow$   & 0.0034 $\pm$ 0.0000 & 0.0034 $\pm$ 0.0000 & 0.0034 $\pm$ 0.0000 & \textbf{0.0032 $\pm$ 0.0001} \\
MAE $\downarrow$   & 0.0165 $\pm$ 0.0001 & 0.0165 $\pm$ 0.0000 & 0.0165 $\pm$ 0.0000 & \textbf{0.0162 $\pm$ 0.0001} \\
GMSD $\downarrow$  & 0.1140 $\pm$ 0.0012 & 0.1121 $\pm$ 0.0002 & 0.1114 $\pm$ 0.0002 & \textbf{0.1097 $\pm$ 0.0004} \\
VIF $\uparrow$     & 0.2379 $\pm$ 0.0031 & 0.2335 $\pm$ 0.0039 & 0.2391 $\pm$ 0.0022 & \textbf{0.2522 $\pm$ 0.0066} \\
FSIM $\uparrow$    & 0.9506 $\pm$ 0.0021 & 0.9578 $\pm$ 0.0002 & 0.9575 $\pm$ 0.0005 & \textbf{0.9581 $\pm$ 0.0003} \\
VSI $\uparrow$     & 0.9326 $\pm$ 0.0004 & \textbf{0.9346 $\pm$ 0.0001} & 0.9343 $\pm$ 0.0003 & 0.9342 $\pm$ 0.0005 \\
HaarPSI $\uparrow$ & 0.4474 $\pm$ 0.0018 & 0.4512 $\pm$ 0.0004 & 0.4536 $\pm$ 0.0008 & \textbf{0.4692 $\pm$ 0.0020} \\
\bottomrule
\end{tabularx}
\end{table*}

The Pix2Pix-to-NSNet-Base comparison was metric dependent. NSNet-Base had higher means for MS-SSIM, PSNR, FSIM, VSI, and HaarPSI and a lower GMSD, whereas Pix2Pix had higher SSIM and VIF; MSE and MAE were unchanged at the reported precision. Thus, the convergence-window results do not support a general SSIM improvement or a uniform benefit from next-scale generation.

Replacing the PatchGAN-style discriminator with TransPatchGAN was associated with small increases from NSNet-Base to NSNet-D in SSIM (0.8883 to 0.8894), MS-SSIM (0.8588 to 0.8593), PSNR (26.8227 to 26.8493), VIF (0.2335 to 0.2391), and HaarPSI (0.4512 to 0.4536), together with a lower GMSD (0.1121 to 0.1114). FSIM (0.9578 to 0.9575) and VSI (0.9346 to 0.9343) were marginally lower, and MSE and MAE were unchanged at the reported precision. 

Compared with NSNet-D, TransNSNet had better mean values in nine of ten metrics, with VSI as the exception. These component-wise differences describe trends from single training runs; they do not by themselves demonstrate statistical or causal superiority of either Transformer component.

\subsection{Fixed-Fraction High-Intensity-Region Analysis}

\begin{figure*}[!t]
\centering
\includegraphics[
  width=\textwidth,
  height=0.72\textheight,
  keepaspectratio
]{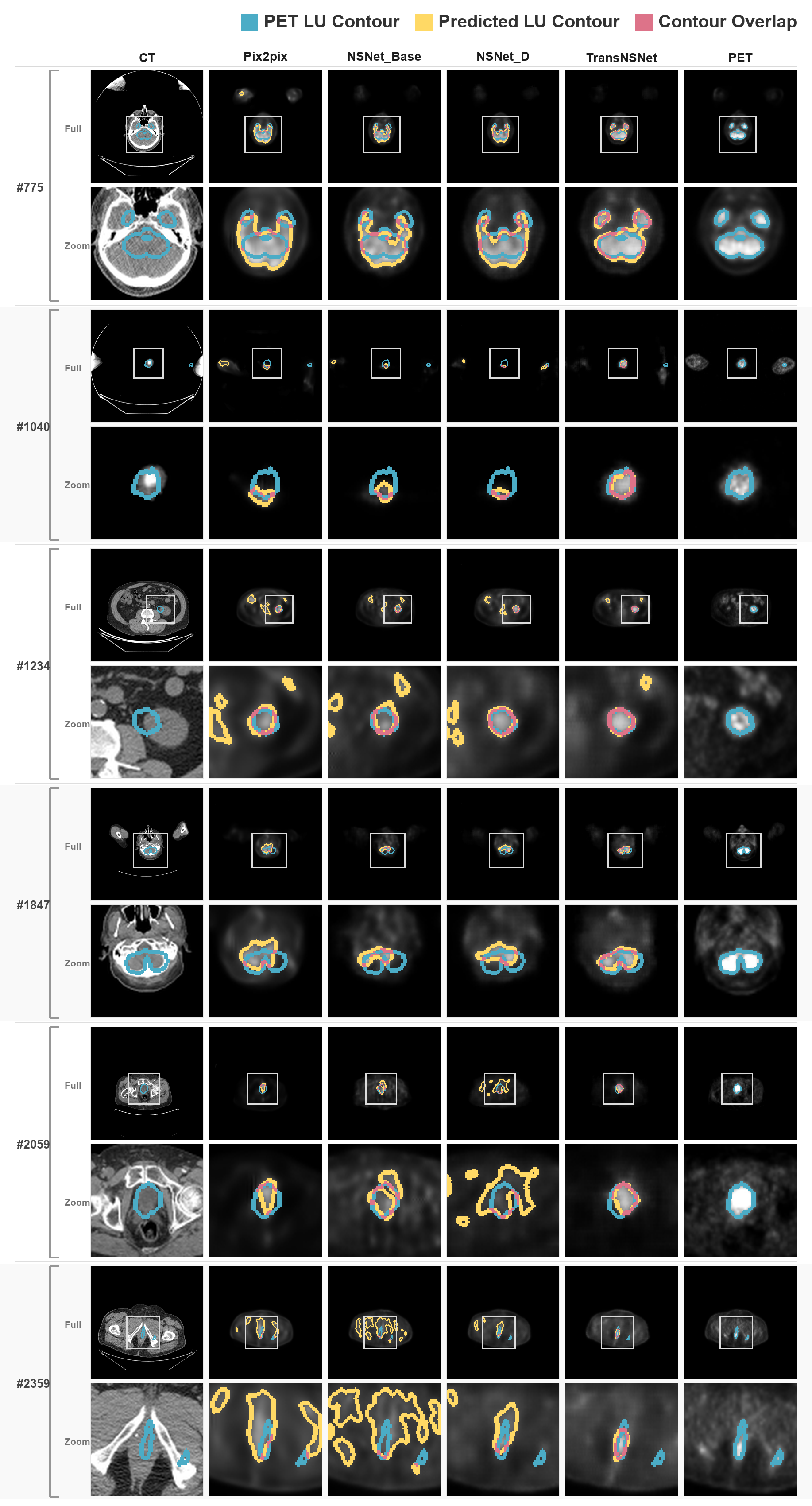}
\caption{Qualitative comparison of adaptive LU contour consistency in selected test cases. Ground-truth and predicted LU masks were independently defined using 50\% of the corresponding slice maximum. Blue, yellow, and pink denote the ground-truth LU contour, predicted LU contour, and their overlap, respectively; the lower row of each case enlarges the boxed region.}
\label{fig:high-uptake-contour}
\end{figure*}

\begin{figure*}[!t]
\centering
\includegraphics[
  width=\textwidth,
  height=0.72\textheight,
  keepaspectratio
]{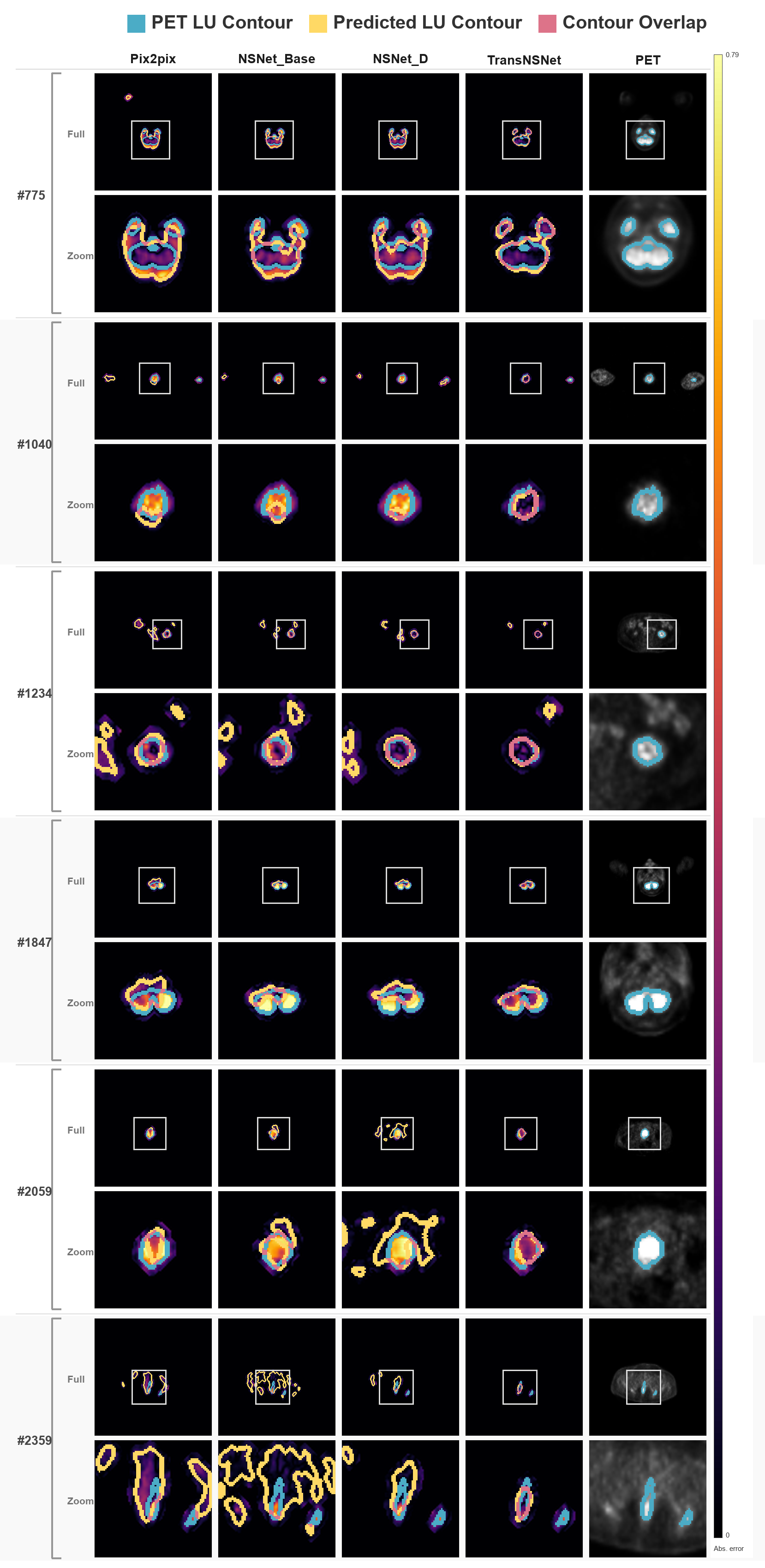}
\caption{Qualitative comparison of absolute-error maps in adaptive LU regions for selected test cases. Ground-truth and predicted LU masks were independently defined using 50\% of the corresponding slice maximum, and errors were calculated as $\lvert I_{\mathrm{pred}}-I_{\mathrm{GT}}\rvert$. Brighter colors indicate larger errors; the lower row of each case enlarges the boxed region, and the rightmost PET column provides the ground-truth reference.}
\label{fig:high-uptake-error}
\end{figure*}

To assess fixed-fraction high-intensity-region consistency, we evaluated the top-5\% masks and related threshold-sensitive measures. Table~\ref{tab:high-uptake} reports these metrics as slice-level mean $\pm$ standard deviation over all 2,905 test slices. The table and Figures~\ref{fig:high-uptake-contour}--\ref{fig:error-maps} use the saved checkpoints specified in the Experimental Settings rather than the epoch-80--101 aggregation used for Table~\ref{tab:main-quantitative}.

\begin{table*}[hbt]
\centering
\caption{Slice-level fixed-fraction high-intensity-region metrics from the saved model-specific checkpoints. Masks are defined by the top-5\% intensity percentile unless otherwise specified. Values are mean $\pm$ slice-level standard deviation over 2,905 slices. The best mean value in each row is shown in bold.}
\label{tab:high-uptake}
\scriptsize
\setlength{\tabcolsep}{2pt}
\renewcommand{\tabularxcolumn}[1]{m{#1}}
\begin{tabularx}{\textwidth}{l*{4}{>{\raggedleft\arraybackslash}X}}
\toprule
Metric & Pix2Pix & NSNet-Base & NSNet-D & TransNSNet \\
\midrule
HU-Dice $\uparrow$ & 0.4063 $\pm$ 0.2187 & 0.4059 $\pm$ 0.2141 & \textbf{0.4230 $\pm$ 0.2156} & 0.4223 $\pm$ 0.2196 \\
HU-IoU $\uparrow$ & 0.2796 $\pm$ 0.1818 & 0.2784 $\pm$ 0.1790 & 0.2931 $\pm$ 0.1844 & \textbf{0.2932 $\pm$ 0.1860} \\
HU-Precision $\uparrow$ & 0.4011 $\pm$ 0.2225 & 0.4032 $\pm$ 0.2183 & 0.4145 $\pm$ 0.2193 & \textbf{0.4165 $\pm$ 0.2244} \\
HU-Recall $\uparrow$ & 0.4176 $\pm$ 0.2187 & 0.4150 $\pm$ 0.2134 & \textbf{0.4402 $\pm$ 0.2171} & 0.4359 $\pm$ 0.2198 \\
HU-Specificity $\uparrow$ & 0.9687 $\pm$ 0.0115 & \textbf{0.9695 $\pm$ 0.0110} & 0.9689 $\pm$ 0.0116 & 0.9693 $\pm$ 0.0118 \\
HU-Balanced Acc. $\uparrow$ & 0.6932 $\pm$ 0.1147 & 0.6922 $\pm$ 0.1118 & \textbf{0.7046 $\pm$ 0.1140} & 0.7026 $\pm$ 0.1154 \\
HU-MAE $\downarrow$ & 0.2425 $\pm$ 0.1354 & 0.2455 $\pm$ 0.1385 & 0.2378 $\pm$ 0.1411 & \textbf{0.2320 $\pm$ 0.1390} \\
HU-RMSE $\downarrow$ & 0.2738 $\pm$ 0.1322 & 0.2757 $\pm$ 0.1346 & 0.2682 $\pm$ 0.1376 & \textbf{0.2616 $\pm$ 0.1363} \\
HU-Peak Error $\downarrow$ & \textbf{0.2655 $\pm$ 0.1189} & 0.2792 $\pm$ 0.1294 & 0.3026 $\pm$ 0.1328 & 0.2715 $\pm$ 0.1323 \\
Centroid Dist. (px) $\downarrow$ & 11.3375 $\pm$ 12.3596 & \textbf{10.7993 $\pm$ 11.5144} & 11.0389 $\pm$ 9.9493 & 11.5538 $\pm$ 10.5895 \\
Peak Loc. Dist. (px) $\downarrow$ & 24.9106 $\pm$ 22.9789 & 24.9236 $\pm$ 23.0581 & \textbf{24.8242 $\pm$ 23.0103} & 25.4830 $\pm$ 22.7863 \\
HD95 (px) $\downarrow$ & 21.8508 $\pm$ 17.3670 & \textbf{21.1215 $\pm$ 16.7420} & 22.3290 $\pm$ 16.4347 & 22.3735 $\pm$ 15.4805 \\
ASSD (px) $\downarrow$ & 7.0408 $\pm$ 7.7261 & \textbf{6.7143 $\pm$ 6.2607} & 6.9521 $\pm$ 5.9753 & 7.0351 $\pm$ 6.0639 \\
False Uptake Area Ratio $\downarrow$ & 0.0313 $\pm$ 0.0115 & \textbf{0.0305 $\pm$ 0.0110} & 0.0311 $\pm$ 0.0116 & 0.0307 $\pm$ 0.0118 \\
False Uptake Burden $\downarrow$ & 0.4770 $\pm$ 0.1407 & \textbf{0.4642 $\pm$ 0.1367} & 0.4658 $\pm$ 0.1419 & 0.4676 $\pm$ 0.1529 \\
Top-3\% Dice $\uparrow$ & 0.3296 $\pm$ 0.2231 & 0.3296 $\pm$ 0.2216 & \textbf{0.3474 $\pm$ 0.2280} & 0.3456 $\pm$ 0.2275 \\
Top-10\% Dice $\uparrow$ & 0.5094 $\pm$ 0.1904 & 0.5075 $\pm$ 0.1870 & 0.5242 $\pm$ 0.1811 & \textbf{0.5257 $\pm$ 0.1921} \\
\bottomrule
\end{tabularx}
\end{table*}

The fixed-fraction high-intensity analysis showed a more nuanced pattern than the image-level metrics, and the relative rankings of NSNet-D and TransNSNet varied slightly with the aggregation unit. At the slice level (Table~\ref{tab:high-uptake}), NSNet-D had marginally higher HU-Recall (0.4402 versus 0.4359) and HU-Balanced Accuracy (0.7046 versus 0.7026), and it also had the highest HU-Dice and Top-3\% Dice. After equal-weight patient-level aggregation (Supplementary Table~\ref{tab:supp-top5-patient}), HU-Recall (0.4397 versus 0.4389) and HU-Balanced Accuracy (0.7044 versus 0.7042) were nearly identical between NSNet-D and TransNSNet, while the highest HU-Dice and Top-3\% Dice shifted to TransNSNet. These differences were small relative to the corresponding between-patient standard deviations and should not be interpreted as clear superiority of either model. TransNSNet had the most favorable numerical values for HU-IoU, HU-Precision, HU-MAE, HU-RMSE, and Top-10\% Dice in the slice-level table, whereas boundary- and location-based results remained mixed across models.

Figures~\ref{fig:high-uptake-contour} and~\ref{fig:high-uptake-error} present slice-adaptive lesion-like uptake (LU) visualizations for selected test cases using a separate 50\%-of-slice-maximum definition, which differs from the top-5\% fixed-fraction HU masks in Table~\ref{tab:high-uptake}.
For each normalized 2D slice, the ground-truth PET image and each predicted image were independently thresholded at 50\% of their respective slice maxima. 
All 4-connected components with an area of at least 20 pixels were retained, and no empty-mask fallback was applied. Figure~\ref{fig:high-uptake-contour} shows contour agreement, whereas Figure~\ref{fig:high-uptake-error} shows absolute-error maps within the union of the ground-truth and predicted high-intensity masks after dilation by five pixels. These threshold-derived visualizations are intended for qualitative technical image-quality assessment and should not be interpreted as radiologist-defined lesion annotations or clinical validation.

\subsection{Patient-Level Adaptive Uptake Analysis}

Table~\ref{tab:adaptive-patient} reports the complementary adaptive lesion-like uptake analysis. Unlike the fixed-fraction slice-level analysis in Table~\ref{tab:high-uptake}, it assigns equal weight to each of the ten patients after averaging evaluable slices within that patient. A total of 2,744 slices were evaluable; 161 slices without an adaptive uptake region were excluded. The LU prefix distinguishes these adaptive metrics from the HU metrics based on top-5\% fixed-fraction masks.

\begin{table*}[htb]
\centering
\caption{Five patient-level adaptive lesion-like uptake metrics. Ground-truth and predicted regions were independently defined at or above 50\% of their respective normalized slice maxima. All 4-connected components containing at least 20 pixels were retained without an empty-mask fallback, and slices were included only when the filtered ground-truth mask contained at least 30 pixels. Values are mean $\pm$ between-patient standard deviation across ten equally weighted patients after within-patient slice averaging (2,744 evaluable slices; 161/2,905 excluded). The best mean value in each row is shown in bold.}
\label{tab:adaptive-patient}
\scriptsize
\setlength{\tabcolsep}{2pt}
\begin{tabularx}{\textwidth}{l*{4}{>{\raggedleft\arraybackslash}X}}
\toprule
Metric & Pix2Pix & NSNet-Base & NSNet-D & TransNSNet \\
\midrule
LU-Dice $\uparrow$ & 0.4232 $\pm$ 0.1267 & 0.4209 $\pm$ 0.1270 & 0.4277 $\pm$ 0.1245 & \textbf{0.4307 $\pm$ 0.1282} \\
LU-IoU $\uparrow$ & 0.3053 $\pm$ 0.1023 & 0.3034 $\pm$ 0.1029 & 0.3108 $\pm$ 0.1016 & \textbf{0.3154 $\pm$ 0.1047} \\
LU-Precision $\uparrow$ & 0.4059 $\pm$ 0.1186 & 0.4101 $\pm$ 0.1089 & 0.4000 $\pm$ 0.1030 & \textbf{0.4155 $\pm$ 0.1015} \\
LU-Recall $\uparrow$ & 0.6215 $\pm$ 0.1747 & 0.6282 $\pm$ 0.1824 & \textbf{0.6953 $\pm$ 0.1790} & 0.6435 $\pm$ 0.2017 \\
LU-MAE $\downarrow$ & 0.2573 $\pm$ 0.0750 & 0.2601 $\pm$ 0.0747 & 0.2500 $\pm$ 0.0779 & \textbf{0.2458 $\pm$ 0.0824} \\
\bottomrule
\end{tabularx}
\end{table*}

The adaptive analysis again showed metric-dependent trends. TransNSNet had the highest patient-level means for LU-Dice, LU-IoU, and LU-Precision and the lowest LU-MAE, whereas NSNet-D had the highest LU-Recall. Thus, TransNSNet had the best mean for four of the five reported metrics. In patient-by-patient comparisons, TransNSNet exceeded Pix2Pix for Dice in 7/10 patients and exceeded NSNet-D for IoU in 8/10 patients. Across the 15 paired comparisons (five metrics and three model comparisons per metric), three had unadjusted $p<0.05$, but none remained significant after Holm correction. The observed differences are therefore descriptive trends rather than evidence of statistical superiority. The limited patient count also means that the absence of significance should not be interpreted as proof of equivalence.

Between-patient variability was substantially larger than the model-level mean differences. For example, the TransNSNet LU-Dice standard deviation was 0.1282, whereas the largest difference among model mean Dice values was approximately 0.010. One patient had Dice values near 0.12 for all four models, further indicating a strong case-specific influence. Supplementary Table~\ref{tab:supp-top5-patient} provides the separate equal-weight patient-level summary for the top-5\% fixed-fraction definition. Because its mask definition and included slices differ, it should not be compared numerically with Table~\ref{tab:adaptive-patient}.

\subsection{Model Parameter Analysis}

To examine whether the favorable image-level trend was associated with a larger model, we further compared the trainable parameter counts recorded in the experiment logs. Table~\ref{tab:model-parameters} summarizes the generator, discriminator, and total parameter counts of the four compared architectures.

\begin{table*}[htb]
\centering
\caption{Model parameter comparison for the four compared architectures. Values are rounded independently to two decimal places in millions from the raw parameter counts; totals were calculated before rounding.}\label{tab:model-parameters}
\scriptsize
\setlength{\tabcolsep}{3pt}
\begin{tabular}{lccc}
\toprule
Model & Generator parameters & Discriminator parameters & Total parameters \\
\midrule
Pix2Pix & 30.94M & 2.77M & 33.70M \\
NSNet-Base & 83.06M & 2.77M & 85.83M \\
NSNet-D & 83.06M & 3.82M & 86.88M \\
TransNSNet & 19.84M & 3.82M & 23.66M \\
\bottomrule
\end{tabular}
\end{table*}

Although TransPatchGAN slightly increases the discriminator parameter count compared with the PatchGAN-style discriminator, the Transformer-enhanced generator used in TransNSNet is substantially more compact than the original next-scale generator. As a result, TransNSNet contains 23.66M parameters in total, which is approximately 72.4\% fewer than NSNet-Base and 72.8\% fewer than NSNet-D. Therefore, the favorable image-level trend for TransNSNet is not associated with a larger parameter count. This analysis is limited to trainable parameter count; inference time, FLOPs, training time, and GPU memory consumption require separate evaluation.

\subsection{Ablation Analysis}

The progressive ablation study was designed for component-wise architectural comparisons. Pix2Pix to NSNet-Base compares the conventional U-Net and next-scale generators; NSNet-Base to NSNet-D compares the PatchGAN-style discriminator and TransPatchGAN; and NSNet-D to TransNSNet compares the original and Transformer-enhanced next-scale generators.

The convergence-window comparison from Pix2Pix to NSNet-Base was not uniformly favorable: NSNet-Base improved PSNR, GMSD, FSIM, VSI, and HaarPSI, but was slightly lower in SSIM and VIF and showed no clear advantage in the fixed-fraction regional metrics. Its contribution is therefore metric dependent rather than a general improvement in structural similarity. The NSNet-Base-to-NSNet-D comparison likewise showed small, mixed image-level differences. For the fixed-fraction regional metrics, NSNet-D was marginally higher than TransNSNet in slice-level HU-Recall and HU-Balanced Accuracy, but their equal-weight patient-level values were nearly identical; the HU-Dice and Top-3\% Dice rankings also shifted from NSNet-D to TransNSNet after patient-level aggregation. From NSNet-D to TransNSNet, nine of ten convergence-window image metrics moved in the favorable direction, while VSI and some regional or localization metrics did not.

These findings are consistent with the design motivation of combining multi-scale prediction and contextual modeling, but the single-run ablation does not establish that any component caused a universal improvement. The evidence instead supports a component-, metric-, and aggregation-dependent interpretation: TransNSNet showed the strongest overall image-level trend, while small regional differences changed rank under slice-level versus patient-level aggregation.

\subsection{Qualitative Evaluation}

\begin{figure*}[h]
\centering
\includegraphics[width=\textwidth,height=0.50\textheight,keepaspectratio]{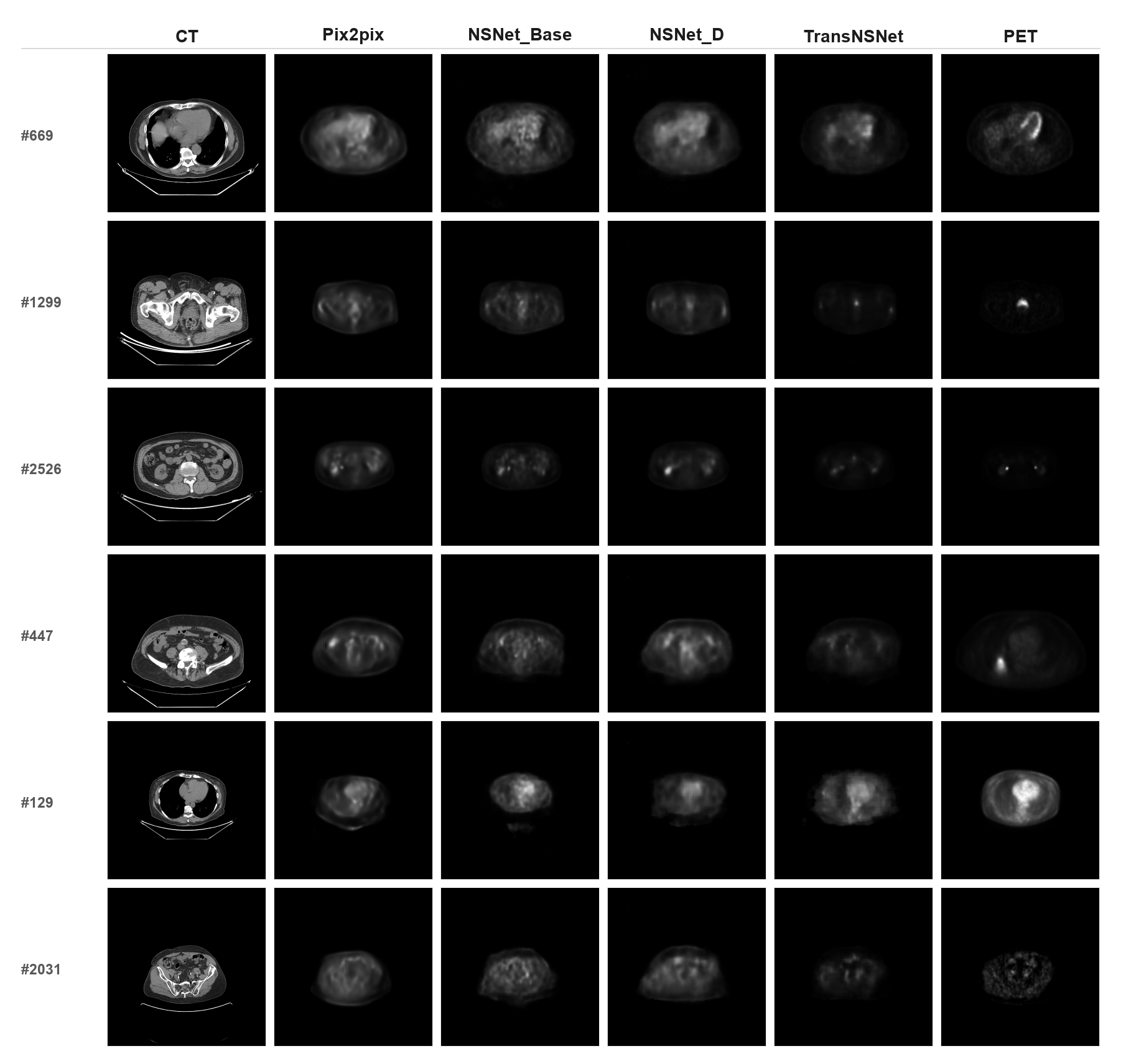}
\caption{Qualitative visual comparison of CT input, Pix2Pix, NSNet-Base, NSNet-D, TransNSNet, and ground-truth PET images using the saved model-specific checkpoints.}
\label{fig:visual-comparison}
\end{figure*}

Qualitative visual comparison was performed using representative CT input images, ground-truth PET images, and synthesized PET images generated by Pix2Pix, NSNet-Base, NSNet-D, and TransNSNet. Figure~\ref{fig:visual-comparison} shows selected axial cases.

In the selected cases, TransNSNet appeared to produce more concentrated PET-like intensity patterns and less diffuse background activation than Pix2Pix and NSNet-Base. In several cases, it also appeared to suppress broad low-contrast artifacts while retaining the main high-intensity pattern. NSNet-D appeared more spatially coherent than NSNet-Base in some slices. These case-specific observations are descriptive and should not be interpreted as evidence of a general performance advantage or a discriminator mechanism.

\begin{figure*}[htb]
\centering
\includegraphics[width=\textwidth,height=0.50\textheight,keepaspectratio]{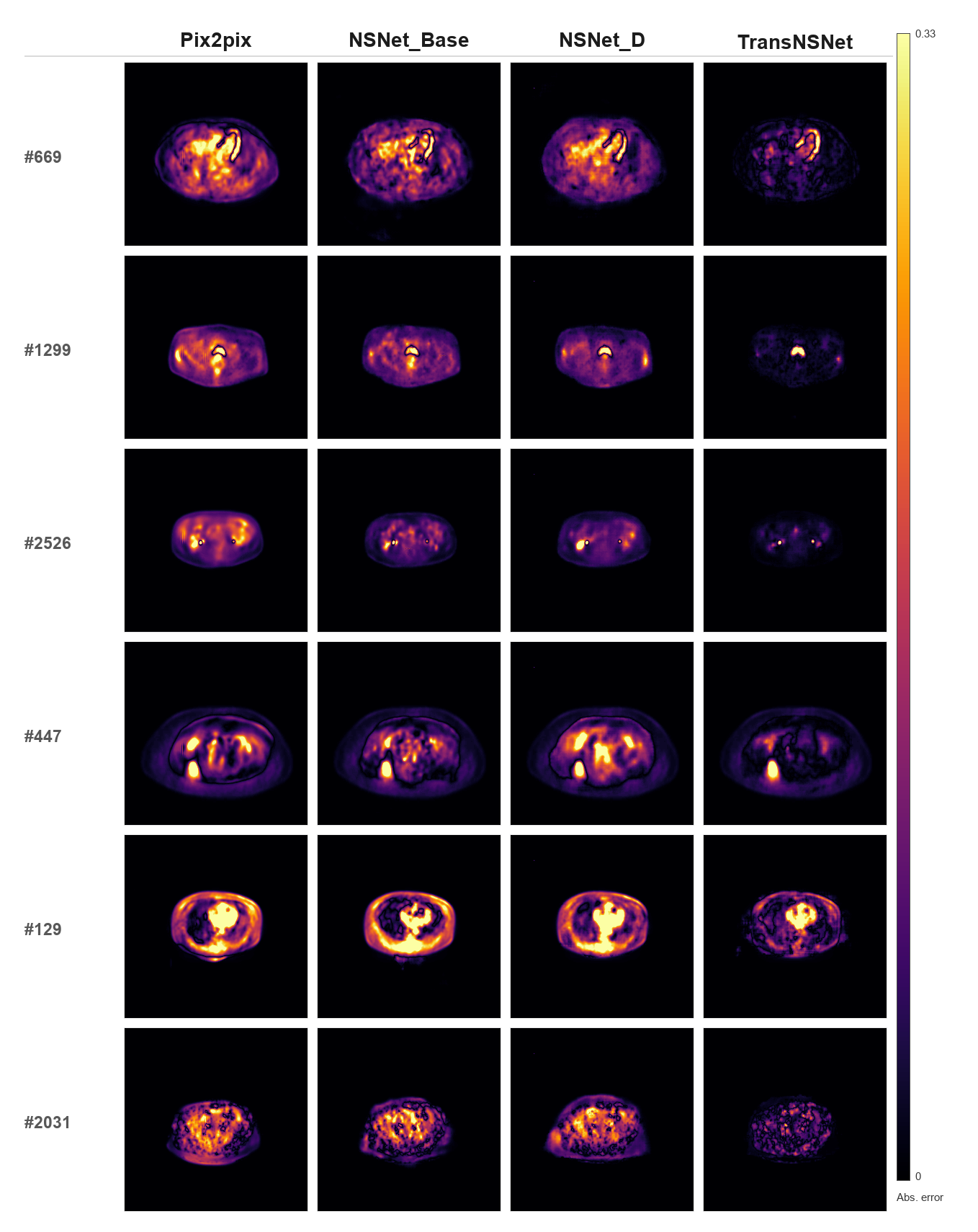}
\caption{Absolute error maps for Pix2Pix, NSNet-Base, NSNet-D, and TransNSNet on selected cases using the saved model-specific checkpoints.}
\label{fig:error-maps}
\end{figure*}

Figure~\ref{fig:error-maps} further presents absolute error maps for the same model set. The TransNSNet error maps show reduced widespread error in several representative slices, especially where the baseline models produce diffuse uptake discrepancies. However, high-intensity boundaries and small focal high-intensity regions remain difficult, and residual errors are still visible in challenging cases. These observations were qualitatively consistent with the quantitative trends while also showing that synthetic PET outputs should be interpreted cautiously.

The 50\%-of-slice-maximum adaptive contour visualization in Figure~\ref{fig:high-uptake-contour} shows where predicted threshold-derived boundaries agree with or deviate from the corresponding PET-derived boundaries in representative cases.

\section{Discussion}

\subsection{Principal Findings}

This study investigated CT-to-PET image synthesis using a Transformer-enhanced next-scale adversarial framework. Over the fixed convergence window, TransNSNet had the best mean value in nine of ten image-level metrics, while NSNet-Base had the highest VSI. The parameter analysis also showed that TransNSNet used fewer trainable parameters than the original next-scale models. These findings indicate a favorable overall image-quality trend for TransNSNet, but not superiority on every measure.

The regional analyses further qualify the image-level result. Under the top-5\% fixed-fraction definition, NSNet-D was marginally higher in slice-level HU-Recall and HU-Balanced Accuracy, whereas the two models were nearly identical for these metrics after equal-weight patient-level aggregation. The highest HU-Dice and Top-3\% Dice also shifted from NSNet-D at the slice level to TransNSNet at the patient level. These aggregation-sensitive differences were small relative to between-patient variability and do not indicate clear superiority of either model; baseline variants also remained best for some boundary and location measures. Under the separate adaptive patient-level definition, TransNSNet had the best mean for four of five metrics, NSNet-D had the highest LU-Recall, and no pairwise difference remained significant after Holm correction.

\subsection{Effect of Next-Scale Generation}

The comparison between Pix2Pix and NSNet-Base examined the replacement of the conventional U-Net generator with next-scale prediction and multi-scale supervision. The convergence-window results were mixed: NSNet-Base improved PSNR, GMSD, FSIM, VSI, and HaarPSI, but its SSIM and VIF were slightly lower than those of Pix2Pix, and the fixed-fraction regional metrics did not show a clear advantage. Consequently, the present data do not support a general claim that next-scale generation improves SSIM or all image-quality measures.

The architectural motivation remains plausible because PET images contain both global uptake patterns and local intensity variations, and supervision at multiple spatial resolutions supplies learning signals at both coarse and fine scales. However, the observed contribution was metric dependent in this single-run study. Moreover, the current design provides multi-scale supervision but does not explicitly feed low-resolution predictions into subsequent higher-resolution stages. Explicit feedback-based coarse-to-fine generation remains a direction for future evaluation.

\subsection{Effect of TransPatchGAN Discriminator}

The convergence-window comparison between NSNet-Base and NSNet-D showed only a limited image-level effect of replacing the PatchGAN-style discriminator with TransPatchGAN. Several structural and error metrics changed in the favorable direction, whereas FSIM and VSI were marginally lower and MSE and MAE were unchanged at the reported precision.

The fixed-fraction regional comparison was sensitive to the aggregation unit. At the slice level, NSNet-D was marginally higher than TransNSNet in HU-Recall and HU-Balanced Accuracy and had the highest HU-Dice and Top-3\% Dice. After equal-weight patient-level aggregation, the two models were nearly identical in HU-Recall and HU-Balanced Accuracy, while the highest HU-Dice and Top-3\% Dice shifted to TransNSNet. Because the differences were small relative to between-patient variability and changed rank with aggregation, they do not provide evidence of a clear regional advantage or verify a discriminator mechanism.

\subsection{Effect of Transformer-Enhanced Generator}

The comparison between NSNet-D and TransNSNet evaluates the additional contribution of introducing Transformer enhancement into the generator. TransNSNet retains the TransPatchGAN discriminator and further replaces the original next-scale generator with a Transformer-enhanced next-scale generator. This design is intended to facilitate long-range feature interaction while preserving multi-scale PET prediction.

This design is motivated by the fact that CT anatomy and PET uptake are not related through purely local visual correspondences. Metabolic activity may be spatially distributed and influenced by broader anatomical and pathological contexts. The present comparison therefore evaluates whether adding generator-side contextual modeling is associated with favorable metric changes; it does not directly verify the proposed long-range mechanism.

In the convergence-window summary, TransNSNet had favorable mean values relative to NSNet-D for nine of ten image-level metrics, with VSI as the exception. The parameter analysis also shows that the Transformer-enhanced generator is more compact than the original next-scale generator. These observations are consistent with a possible generator-side contribution, but the single-run design and mixed regional results do not establish a universal or statistically significant advantage. The result is therefore interpreted as a promising efficiency and performance trend that requires repeated-seed and external validation.

\subsection{Use Boundaries and Technical Interpretation}

The synthesized PET-like images generated by TransNSNet may provide auxiliary PET-like intensity representations derived from CT images for research applications and multimodal image analysis. However, the generated images should not be interpreted as substitutes for clinically acquired PET scans, and the present study does not provide clinical validation of the synthesized outputs.

This limitation is especially important because PET imaging reflects radiotracer uptake and physiological processes that cannot be fully inferred from CT anatomy alone. Although TransNSNet showed favorable mean values for most image-similarity metrics, it does not guarantee accurate recovery of true metabolic activity at the lesion level. Therefore, the current study should be interpreted as a technical investigation of CT-to-PET image synthesis based on image-quality metrics, rather than evidence for diagnostic replacement or clinical use.

Future work should include lesion-level evaluation, SUV-related quantitative analysis if available, and blinded expert reader studies before any clinical use is considered. These analyses are outside the scope of the present study.

\subsection{Limitations}

This study has several limitations. First, the dataset was derived from a single public NSCLC imaging collection, and the test set contained only ten patients. Although the number of slices was large, patients contributed unequal numbers of slices. The primary patient-level analyses therefore averaged within each patient and assigned equal weight to all ten patients. Validation on larger cohorts, more patients, and independent external datasets is required.

Second, all four models were represented by a single training run, without repetition across different random seeds. The standard deviations in Table~\ref{tab:main-quantitative} quantify epoch-to-epoch fluctuation within epochs 80--101 and do not estimate initialization or between-run uncertainty. The small differences among several model means should therefore be regarded as trends until repeated-seed experiments are available.

Third, the adaptive patient-level analysis did not identify any model difference that remained significant after Holm correction. Between-patient variability was large relative to the model means: the TransNSNet LU-Dice standard deviation was 0.128, whereas the largest difference in mean Dice among models was approximately 0.010. One patient had low Dice values near 0.12 for every model. With $n=10$, the analysis also has limited power, so non-significance neither demonstrates superiority nor proves equivalence.

Fourth, parts of the experiments were run on different computers. The same data and experimental settings were used, but unmeasured environment differences may remain. NSNet-Base was interrupted once by a system restart after epoch 87, resumed from saved weights, and had its two evaluation CSV segments manually concatenated without changing values. The other models were not interrupted or manually merged. Reproducing all runs in a controlled environment would further reduce this source of uncertainty.

Fifth, the primary Table~\ref{tab:main-quantitative} uses a fixed convergence window and does not select a single epoch by test performance. However, the saved checkpoints used for Table~\ref{tab:high-uptake} and Figures~\ref{fig:high-uptake-contour}--\ref{fig:error-maps} were retained under the original best-SSIM monitoring rule, without a separate validation-only subset. Those checkpoint-based regional and qualitative findings should therefore remain descriptive. Future work should separate validation-based model selection from final test reporting or use cross-validation.

Sixth, image-level similarity and threshold-derived uptake metrics do not establish true lesion-level metabolic correctness. The adaptive regions are algorithmically defined from normalized intensity rather than delineated by radiologists, and normalized PET-like outputs are not quantitative SUV estimates. Lesion-level annotations, SUV-related analysis where available, blinded reader studies, and task- or outcome-based validation are needed before clinical interpretation.

Finally, the model is two-dimensional and does not explicitly model volumetric continuity. The next-scale mechanism provides multi-scale supervision but not explicit prediction feedback between resolutions. In addition, parameter count alone does not characterize computational cost. Future work should evaluate 3D or hybrid designs, feedback-based coarse-to-fine generation, FLOPs, inference time, training time, and memory consumption.

\section{Conclusion}

This study proposed TransNSNet, a Transformer-enhanced next-scale adversarial framework for CT-to-PET image synthesis, and evaluated it through progressive ablation against Pix2Pix, NSNet-Base, and NSNet-D. Over epochs 80--101, TransNSNet achieved the best mean value in nine of ten image-level metrics, while NSNet-Base achieved the highest VSI. TransNSNet also used substantially fewer trainable parameters than the original next-scale models.

The high-intensity analyses revealed trade-offs that were not captured by the aggregate image metrics. For the top-5\% fixed-fraction definition, NSNet-D was marginally higher in slice-level HU-Recall and HU-Balanced Accuracy, but the patient-level values were nearly identical, and the highest HU-Dice and Top-3\% Dice shifted to TransNSNet after equal-weight patient aggregation. These small, aggregation-dependent differences should not be interpreted as clear superiority. In the separate adaptive patient-level comparison, no model difference remained significant after Holm correction. The present evidence therefore indicates promising performance trends rather than comprehensive or statistically confirmed superiority.

Synthesized PET-like images should be regarded as auxiliary computational outputs rather than substitutes for clinically acquired PET. Larger multi-patient cohorts, repeated random-seed experiments, independent external validation, radiologist- or outcome-based assessment, and three-dimensional modeling are needed to determine robustness and potential utility.

\begingroup
\setcounter{table}{0}
\renewcommand{\thetable}{S\arabic{table}}
\renewcommand{\theHtable}{S\arabic{table}}
\section*{Supplementary Material}

Supplementary Table~\ref{tab:supp-top5-patient} complements the slice-level fixed-fraction results in Table~\ref{tab:high-uptake} by assigning equal weight to each test patient. Under this aggregation, NSNet-D and TransNSNet were nearly identical in HU-Recall and HU-Balanced Accuracy, while HU-Dice and Top-3\% Dice were marginally higher for TransNSNet. It is distinct from the adaptive lesion-like uptake analysis in Table~\ref{tab:adaptive-patient}; the two analyses use different mask definitions and should not be compared directly.

\begin{table*}[htb]
\centering
\caption{Patient-level evaluation of fixed-fraction high-intensity regions defined by the top-5\% intensity percentile. Values are mean $\pm$ between-patient standard deviation across ten patients. Each metric was first averaged over all slices within each patient, after which patient-level means were averaged with equal weights. All 2,905 slices were included.}
\label{tab:supp-top5-patient}
\tiny
\setlength{\tabcolsep}{1.5pt}
\renewcommand{\arraystretch}{1.08}
\begin{tabularx}{\textwidth}{l *{4}{>{\raggedleft\arraybackslash}X}}
\toprule
Metric & Pix2Pix & NSNet-Base & NSNet-D & TransNSNet \\
\midrule
HU-Dice $\uparrow$ & 0.4114 $\pm$ 0.0983 & 0.4103 $\pm$ 0.1009 & 0.4261 $\pm$ 0.0991 & 0.4281 $\pm$ 0.1000 \\
HU-IoU $\uparrow$ & 0.2837 $\pm$ 0.0775 & 0.2817 $\pm$ 0.0798 & 0.2958 $\pm$ 0.0810 & 0.2978 $\pm$ 0.0806 \\
HU-Precision $\uparrow$ & 0.4075 $\pm$ 0.0959 & 0.4091 $\pm$ 0.0985 & 0.4195 $\pm$ 0.0951 & 0.4239 $\pm$ 0.0980 \\
HU-Recall $\uparrow$ & 0.4206 $\pm$ 0.0982 & 0.4169 $\pm$ 0.1001 & 0.4397 $\pm$ 0.1002 & 0.4389 $\pm$ 0.0991 \\
HU-Specificity $\uparrow$ & 0.9689 $\pm$ 0.0050 & 0.9696 $\pm$ 0.0051 & 0.9691 $\pm$ 0.0051 & 0.9696 $\pm$ 0.0053 \\
HU-Balanced Acc. $\uparrow$ & 0.6948 $\pm$ 0.0515 & 0.6932 $\pm$ 0.0526 & 0.7044 $\pm$ 0.0526 & 0.7042 $\pm$ 0.0522 \\
HU-MAE $\downarrow$ & 0.2474 $\pm$ 0.0713 & 0.2498 $\pm$ 0.0702 & 0.2426 $\pm$ 0.0734 & 0.2371 $\pm$ 0.0719 \\
HU-RMSE $\downarrow$ & 0.2788 $\pm$ 0.0705 & 0.2799 $\pm$ 0.0700 & 0.2729 $\pm$ 0.0733 & 0.2666 $\pm$ 0.0719 \\
HU-Peak Error $\downarrow$ & 0.2648 $\pm$ 0.0223 & 0.2789 $\pm$ 0.0227 & 0.3022 $\pm$ 0.0302 & 0.2728 $\pm$ 0.0319 \\
Centroid Dist. (px) $\downarrow$ & 11.0949 $\pm$ 3.9584 & 10.5848 $\pm$ 3.6983 & 10.9709 $\pm$ 3.0845 & 11.3618 $\pm$ 3.6385 \\
Peak Loc. Dist. (px) $\downarrow$ & 24.8531 $\pm$ 6.0030 & 24.8010 $\pm$ 6.4855 & 24.9034 $\pm$ 6.4388 & 25.4348 $\pm$ 5.8154 \\
HD95 (px) $\downarrow$ & 21.8155 $\pm$ 6.6262 & 21.1238 $\pm$ 6.5844 & 22.6192 $\pm$ 6.9257 & 22.4166 $\pm$ 6.4492 \\
ASSD (px) $\downarrow$ & 6.9792 $\pm$ 3.0061 & 6.7134 $\pm$ 2.8591 & 7.0163 $\pm$ 2.8692 & 7.0197 $\pm$ 2.9130 \\
False Uptake Area Ratio $\downarrow$ & 0.0311 $\pm$ 0.0050 & 0.0304 $\pm$ 0.0051 & 0.0309 $\pm$ 0.0051 & 0.0304 $\pm$ 0.0053 \\
False Uptake Burden $\downarrow$ & 0.4739 $\pm$ 0.0380 & 0.4611 $\pm$ 0.0438 & 0.4634 $\pm$ 0.0389 & 0.4639 $\pm$ 0.0697 \\
Top-3\% Dice $\uparrow$ & 0.3339 $\pm$ 0.0914 & 0.3337 $\pm$ 0.0973 & 0.3490 $\pm$ 0.0945 & 0.3500 $\pm$ 0.0957 \\
Top-10\% Dice $\uparrow$ & 0.5142 $\pm$ 0.0997 & 0.5126 $\pm$ 0.0999 & 0.5286 $\pm$ 0.0989 & 0.5314 $\pm$ 0.1034 \\
\bottomrule
\end{tabularx}
\end{table*}
\endgroup

\section*{Data Availability}

The original imaging data used in this study are publicly available from The Cancer Imaging Archive NSCLC Radiogenomics collection and can be accessed through the original dataset record at \url{https://doi.org/10.7937/K9/TCIA.2017.7hs46erv}. The original imaging data and the processed paired CT-PET dataset will not be redistributed by the authors; users should obtain the source data directly from the original dataset website and comply with its data-use conditions. The experimental code builds on a third-party research codebase and is not publicly released owing to code-provenance and laboratory data-management policies.

\section*{Ethics Statement}

This study used publicly available de-identified data from The Cancer Imaging Archive NSCLC Radiogenomics collection. No new human-subject data were collected for this study. The dataset was used in accordance with the data access and usage conditions of The Cancer Imaging Archive.

\section*{AI Assistance Disclosure}
OpenAI GPT-5.5, GPT-5.6-Sol, and GPT-6-Astra accessed through the web-based ChatGPT interface and the desktop Codex application, were used to assist with English language editing and polishing, experimental engineering code adjustment, and figure-generation code editing and engineering. The authors reviewed, edited, and approved all AI-assisted outputs and take full responsibility for the content, experiments, and conclusions of this work.

\section*{Funding}

This research received no specific grant from any funding agency in the public, commercial, or not-for-profit sectors.

\section*{Conflicts of Interest}

The authors declare no conflicts of interest.

\section*{CRediT authorship contribution statement}

\textbf{Shaojun Li:} Conceptualization, Methodology, Software, Data curation, Investigation, Formal analysis, Validation, Visualization, Writing -- original draft, Writing -- review \& editing.

\textbf{Xiaoyu Deng:} Methodology, Software, Supervision, Writing -- review \& editing.

\textbf{Ikuo Keshi:} Methodology, Validation, Supervision, Writing -- review \& editing.

\section*{Acknowledgment}

The authors thank The Cancer Imaging Archive and the contributors of the NSCLC Radiogenomics collection for providing the public dataset used in this study. The authors also thank Fukui University of Technology for providing the research environment, equipment, and support for this work.

\bibliographystyle{model2-names}

\bibliography{refs_list}

\end{document}